\documentclass[12pt]{article}
\usepackage{a4wide,epsfig,psfrag,amsmath,amssymb,cite,scalefnt}
\usepackage{color}

\graphicspath{ {figs/} }

\newcommand\eps{\epsilon}

\newcommand{\ep}{\epsilon}
\newcommand{\nc}{N_c}
\newcommand{\nl}{n_l}
\newcommand{\logx}{l_x}

\allowdisplaybreaks

\begin{document}

\title{\vskip-3cm{\baselineskip14pt
    \begin{flushleft}
     \normalsize MITP/16-110\\
     \normalsize TTP16-053
    \end{flushleft}} \vskip1.5cm 
  Massive three-loop form factor in the planar limit
  }

\author{
  Johannes Henn$^{a}$,
  Alexander V. Smirnov$^{b}$,
  \\
  Vladimir A. Smirnov$^{c,d}$,
  Matthias Steinhauser$^{d}$,
  \\[1em]
  {\small\it (a) PRISMA Cluster of Excellence, Johannes Gutenberg University,}\\
  {\small\it  55099 Mainz, Germany}
  \\
  {\small\it (b) Research Computing Center, Moscow State University}\\
  {\small\it 119991, Moscow, Russia}
  \\  
  {\small\it (c) Skobeltsyn Institute of Nuclear Physics of Moscow State University}\\
  {\small\it 119991, Moscow, Russia}
  \\
  {\small\it (d) Institut f{\"u}r Theoretische Teilchenphysik,
    Karlsruhe Institute of Technology (KIT)}\\
  {\small\it 76128 Karlsruhe, Germany}  
}
  
\date{}

\maketitle

\thispagestyle{empty}

\begin{abstract}

  We compute the three-loop QCD corrections to the massive quark-anti-quark-photon
  form factors $F_1$ and $F_2$ in the large-$N_c$ limit.  The analytic results are
  expressed in terms of Goncharov polylogarithms. This allows for a straightforward
  numerical evaluation. We also derive series expansions, including power suppressed terms, 
  for three kinematic regions corresponding to small and large invariant masses of the photon
  momentum, and small velocities of the heavy quarks. 

  \medskip

  \noindent
  PACS numbers: 12.38.Bx, 12.38.Cy, 11.15.Bt, 14.65.Ha

\end{abstract}

\thispagestyle{empty}


\newpage



\section{Introduction}

Massive form factors are important building blocks for various physical
quantities involving heavy quarks.
Among them are static quantities like anomalous magnetic moments
but also production cross sections and decay rates.
Furthermore, form factors are the prime examples for
studying the infrared behaviour of QCD amplitudes.

We consider QCD corrections to the quark-photon vertex. The latter can be
parametrized as follows,
\begin{eqnarray}
  V^\mu(q_1,q_2) &=& \bar{u}(q_1) \Gamma^\mu(q_1,q_2) v(q_2)
  \,,
  \label{eq::Vmu}
\end{eqnarray}
where the colour indices of the quarks are suppressed and
$\bar{u}(q_1)$ and $v(q_2)$ are the spinors of the quark and anti-quark,
respectively.  The momentum $q_1$ is incoming and $q_2$ is outgoing with
$q_1^2=q_2^2=m^2$.

The vertex function $\Gamma^\mu(q_1,q_2)$ can be decomposed into
two scalar form factors which are usually introduced as
\begin{eqnarray}
  \Gamma^\mu(q_1,q_2) &=& Q_q
  \left[F_1(q^2)\gamma^\mu - \frac{i}{2m}F_2(q^2) \sigma^{\mu\nu}q_\nu\right]
  \,,
  \label{eq::Gamma}
\end{eqnarray}
where $q=q_1-q_2$ is the outgoing momentum of the photon and
$\sigma^{\mu\nu} = i[\gamma^\mu,\gamma^\nu]/2$.
$Q_q$ is the charge of the considered quark.
$F_1$ and $F_2$ are often referred to as electric and magnetic form factors.

\begin{figure}[b] 
  \begin{center}
    \includegraphics[width=0.2\textwidth]{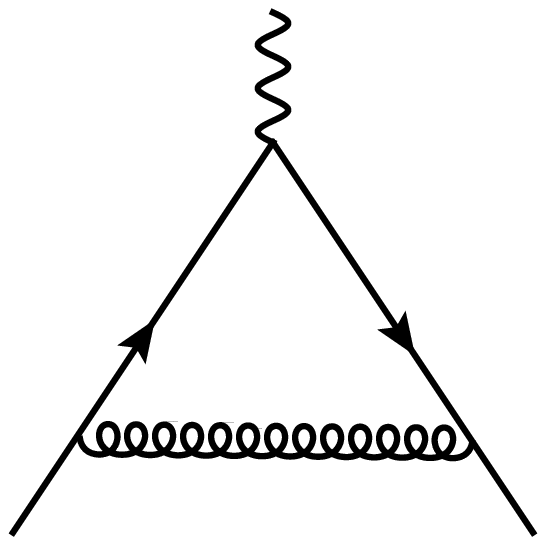}  \hfill
    \includegraphics[width=0.2\textwidth]{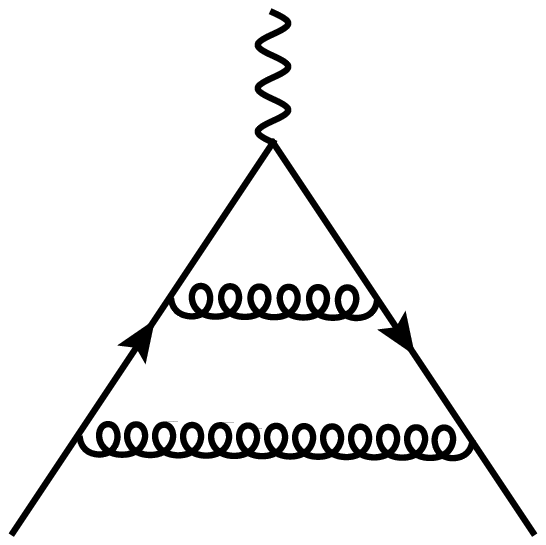}\hfill
    \includegraphics[width=0.2\textwidth]{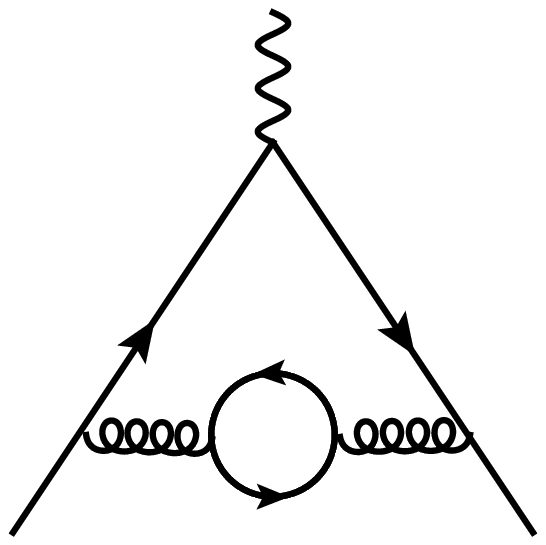}\hfill
    \includegraphics[width=0.2\textwidth]{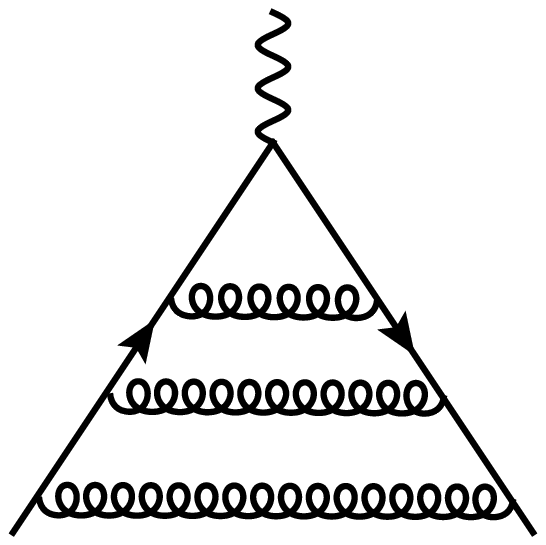}
    \\
    \includegraphics[width=0.2\textwidth]{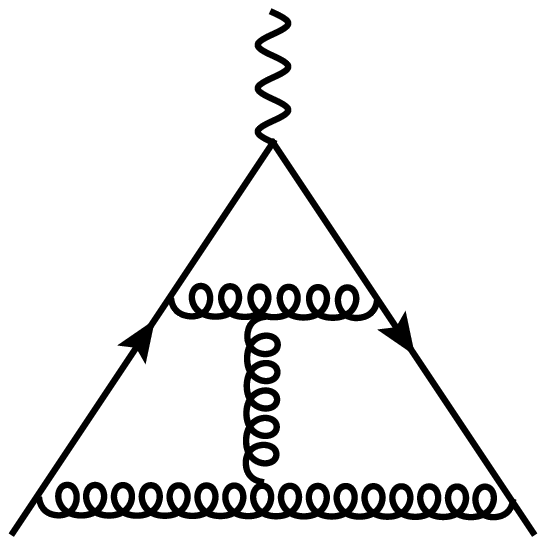}\hfill
    \includegraphics[width=0.2\textwidth]{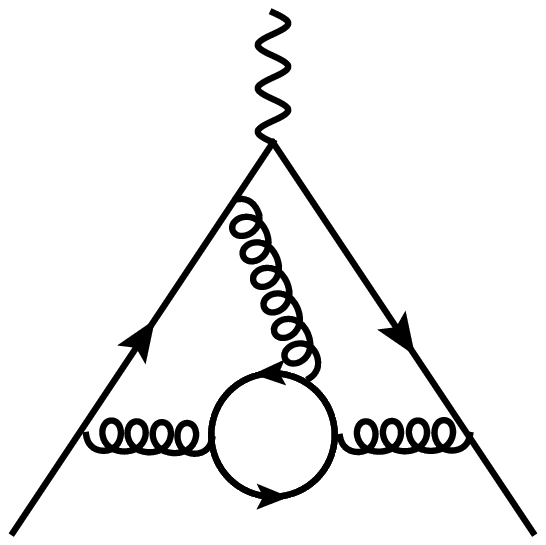}\hfill
    \includegraphics[width=0.2\textwidth]{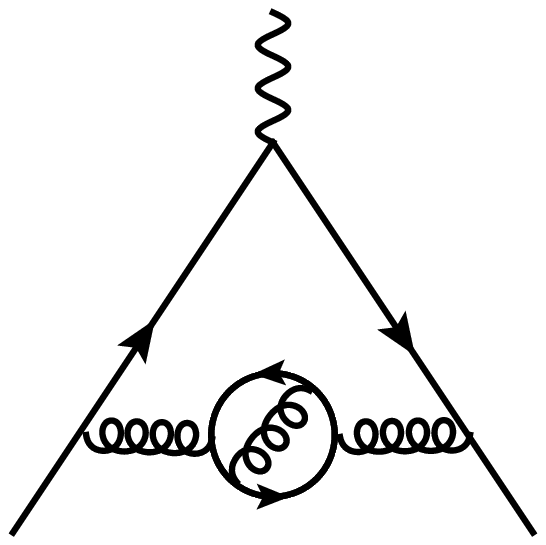}\hfill
    \includegraphics[width=0.2\textwidth]{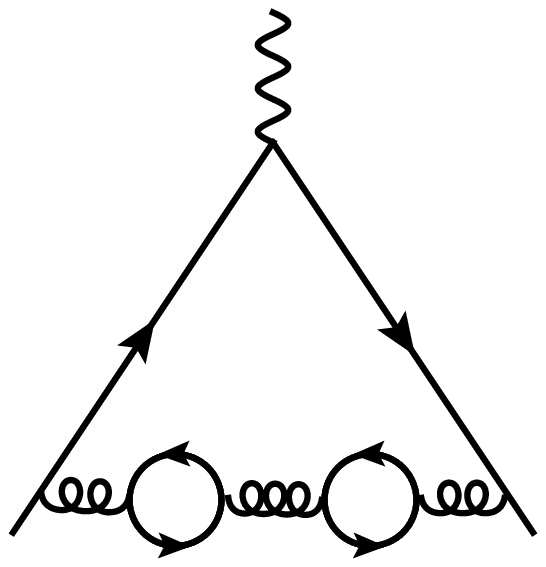}
    \caption{\label{fig::diags}Sample diagrams contributing to $F_1$ and $F_2$
    at one-, two- and three-loop order. Solid, curly and wavy lines represent
    quarks, gluons and photons, respectively. In our calculation the closed
    fermion loops only involve massless quarks.}
  \end{center}
\end{figure}

Sample Feynman diagrams can be found in Fig.~\ref{fig::diags}.  Two-loop QCD
corrections to the electric and magnetic form factors for the heavy quark
vector current have been computed for the first time in
Ref.~\cite{Bernreuther:2004ih} (axial vector and anomaly contributions have
been considered in~\cite{Bernreuther:2004th,Bernreuther:2005rw}) where
analytic results have been obtained.  An independent cross check of the
two-loop results for $F_1$ and $F_2$ has been performed in~\cite{Gluza:2009yy}
where also ${\cal O}(\epsilon^2)$ and ${\cal O}(\epsilon)$ terms have been
added to the one- and two-loop results, respectively.  The results have been
used to obtain predictions for the three-loop form factor $F_1$ in the high
energy limit, by exploiting evolution equations and the exponentiation of
infrared divergences (see also Ref.~\cite{Mitov:2006xs} for earlier
considerations).

In this paper we compute the three-loop form factor in the planar limit,
keeping the exact mass dependence.  After expanding our exact result for small
quark masses we can compare to the high-energy results of~\cite{Gluza:2009yy}
mentioned above, and complete them by determining the unknown constants in the
$1/\epsilon$ and $\epsilon^0$ part.  We furthermore provide power-suppressed
terms.

Massive form factors have infrared divergences that are well understood.
After the ultraviolet renormalization, all poles in dimensional regularization
are given in terms of the cusp anomalous
dimension~\cite{Polyakov:1980ca,Korchemsky:1987wg}, and the beta function.
The three-loop cusp anomalous dimension was computed in
Ref.~\cite{Grozin:2014hna,Grozin:2015kna}.  By verifying the infrared pole
structure at the three-loop order, we provide a first independent check of the
result of Ref.~\cite{Grozin:2014hna,Grozin:2015kna} (in the planar limit).

In the static limit, the infrared divergences disappear, and $F_1$ and $F_2$ are finite.
In fact, $F_1$ vanishes and $F_2$ determines the anomalous magnetic
moment of a heavy quark which has been considered at two-loop order
in Ref.~\cite{Bernreuther:2005gq}. A dedicated calculation at three loops
has been performed in Ref.~\cite{Grozin:2007fh} which serves as a welcome
check for our exact result expanded for $q^2\to0$.

The remainder of the paper is organized as follows.  In
Section~\ref{sec::calc} we provide technical details on the calculation of the
amplitudes. In particular we briefly describe the renormalization procedure.
The infrared structure of the form factors is presented in
Section~\ref{sec::IR}.  Our results for $F_1$ and $F_2$ are discussed in
Section~\ref{sec::res} including the three-loop results for the
static limit, the high-energy limit and for small quark velocities.  We
conclude in Section~\ref{sec::concl}.


\section{\label{sec::calc}Setup and calculation}

The form factors $F_1$ and $F_2$ appearing in Eq.~(\ref{eq::Gamma})
are conveniently computed with the help
of projectors which are applied to $\Gamma^\mu(q_1,q_2)$. Using the kinematics
defined in Eq.~(\ref{eq::Gamma}) we have ($i=1,2$)\footnote{Note that there is
  a typo in Eq.~(11) of~\cite{Bernreuther:2004ih}: $[4/s-2+2\epsilon]$ should
  read $[4/s+2-2\epsilon]$.}
\begin{eqnarray}
  F_i &=& \frac{1}{Q_q}
  \mbox{Tr}\left\{
  (q_1\!\!\!\!\!/\,\,\,+m) 
  \left[ 
   a_{F_i} \gamma_\mu + b_{F_i} \frac{(q_{1,\mu} + q_{2,\mu})}{2m}
  \right]
 (q_2\!\!\!\!\!/\,\,\,+m)
  \Gamma^\mu(q_1,q_2)
  \right\}
  \,,
\end{eqnarray}
with
\begin{eqnarray}
  a_{F_1} = \frac{1}{4(1-\epsilon)(s-4m^2)}\,,&&
  b_{F_1} = \frac{(3 - 2\epsilon) m^2}{(1 - \epsilon)(s-4m^2)^2}\,,
  \nonumber\\
  a_{F_2} = -\frac{m^2}{(1-\epsilon)s(s-4m^2)}\,,&&
  b_{F_2} = -\frac{2m^2(2m^2+s-s\epsilon)}{(1 - \epsilon)s(s-4m^2)^2}\,,
\end{eqnarray}
and $s=q^2$. It is convenient to introduce the dimensionless
variable
\begin{eqnarray}
  \frac{s}{m^2} &=& - \frac{(1-x)^2}{x}
  \,.
\end{eqnarray}
Then the low-energy, high-energy and threshold limits
correspond to $x\to 1$, $x\to 0$ and $x\to -1$, respectively.
Note that for $x>0$ we have $s<0$ and thus the form factors do
not have imaginary parts. The same is true for $x\in \mathbb{C}$ 
with $|x|=1$. For $0<s<4m^2$ we have that $x$ is on the upper
half of the unit circle.

It is convenient to write the perturbative expansion of $F_i$ ($i=1,2$)
in the form
\begin{eqnarray}
  F_i &=& \sum_{n\ge0} \left(\frac{\alpha_s}{4\pi}\right)^n F_i^{(n)}(x)
  \,,
  \label{eq::F_i}
\end{eqnarray}
with $F_1^{(0)}=1$ and $F_2^{(0)}=0$.
In the large-$N_c$ limit we furthermore have that
$F_i^{(1)}\sim N_c$, 
$F_i^{(2)}\sim N_c^2, N_c n_l$, and
$F_i^{(3)}\sim N_c^3, N_c^2 n_l, N_c n_l^2$,
where $n_l$ counts the number of closed massless quark loops
and $N_c$ is the number of colours. Note that we do not consider contributions
with massive closed fermion loops.
In Eq.~(\ref{eq::F_i}) we suppress the scale dependence of $\alpha_s$
and $F_i^{(n)}$.

The calculations performed in this paper use the groundwork performed in
Ref.~\cite{HSS16} where all scalar integral families up to three loops, which
are needed for the massive form factors $F_1$ and $F_2$ in the large-$N_c$
limit, have been classified and the corresponding master integrals have been
computed analytically in terms of Goncharov
polylogarithms~\cite{Goncharov:1998kja}.  We use in particular the information
from Fig.~1 of Ref.~\cite{HSS16} where eight three-loop families are defined.
This information is used to generate with the help of the programs {\tt
  qgraf}~\cite{Nogueira:1991ex} and {\tt
  q2e/exp}~\cite{Harlander:1997zb,Seidensticker:1999bb} amplitudes for $F_1$
and $F_2$ which are expressed in terms of linear combinations of
integrals from the eight three-loop families.
We also use formulae for reduction of Goncharov polylogarithm values at
sixth roots of unity derived in~\cite{Henn:2015sem}.

For the reduction to master integrals we use the program {\tt
  FIRE}~\cite{Smirnov:2008iw,Smirnov:2013dia,Smirnov:2014hma} in combination
with {\tt LiteRed}~\cite{Lee:2012cn,Lee:2013mka}.  Once the reduction for each
family is complete we use the program {\tt tsort}, which is part of the latest
{\tt FIRE} version~\cite{Smirnov:2014hma} and based on ideas presented in
Ref.~\cite{Smirnov:2013dia}, to obtain relations between primary master
integrals, and to arrive at a minimal set.  This leads to 89 master integrals
needed for the large-$N_c$ limit of $F_1$ and $F_2$.

In our calculation we allow for a general QCD gauge parameter $\xi$ but set
$\xi^2$ terms ($\xi=0$ corresponds to Feynman gauge) to zero before performing
the reduction to master integrals. The bare form factors still contain linear
$\xi$ terms which only drop out after renormalization. This serves as a welcome
check for our calculation.

The ultraviolet renormalized form factors are obtained by renormalizing the strong
counpling constant $\alpha_s$ in the $\overline{\rm MS}$ scheme and the heavy
quark mass on-shell.  Both counterterms are needed to two-loop accuracy and
are well-known in the literature. Note, however, that for the on-shell mass
counterterm higher order $\epsilon$ terms are needed. The latter can be found in
Ref.~\cite{Marquard:2007uj}.

In this context we would like to mention that in
Ref.~\cite{Bernreuther:2004ih} an non-standard version of the $\overline{\rm
  MS}$ scheme has been employed as can bee seen from Eq.~(24) of that
paper. The quantity $C(\epsilon)$, which enters the definition of the
renormalization constant, induces $\pi^2$ terms which enter the $\epsilon^0$
part of the two-loop form factor.  See also the discussion in
Ref.~\cite{Mitov:2006xs} on this subject.

A further ingredient to the renormalization procedure is the
on-shell wave function renormalization constant
for the external heavy quarks which is needed to three-loop order
and can be found in Refs.~\cite{Melnikov:2000zc,Marquard:2007uj}.


\section{\label{sec::IR}Infrared divergences of massive form factors}

Form factors of massive particles have infrared divergences originating from
exchanges of soft particles. The latter can be described in the eikonal
approximation. In this way, the infrared divergences of the form factors can
be mapped to ultraviolet divergences of Wilson lines~\cite{Korchemsky:1991zp}.
The relevant Wilson line has the geometry of a cusp formed by the particle
momenta.  It obeys a renormalization group equation that is governed by the
cusp anomalous
dimension~\cite{Polyakov:1980ca,Brandt:1981kf,Korchemsky:1987wg}.

Applying this correspondence to the original form factors, one has
\begin{align}\label{ff_ir_ren}
  F = Z F^{f}\,,
\end{align}
where $Z$ is an infrared renormalization factor (in minimal subtraction), $F$
is the ultraviolet-renormalized form factor, and $F^{f}$ is finite both in the
ultraviolet and infrared.  In other words, all infrared poles of $F$ are
reproduced by $Z$.

$Z$ satisfies the following renormalization group equation 
\begin{align}\label{rg_eq_ir}
  \left[2 \beta_D({\alpha}_s,\eps) {\alpha}_s
    \frac{\partial}{\partial{{\alpha}_s} }+ \Gamma_{\rm cusp}(\phi
    ,{\alpha}_s,\eps) \right] Z({\alpha}_s,\eps) = 0
  \,,
\end{align}
where ${\alpha}_s$ is the renormalized strong coupling and
$\beta_D$ is the $D$-dimensional $\beta$ function,
\begin{align}
  \beta_{D} =  \eps + \sum_{i \ge 1} \beta_{i-1} \left(\frac{{\alpha}_s}{4 \pi}\right)^i  \,,
\end{align}
with
\begin{eqnarray}
  \beta_0 &=& \frac{11}{3}C_A - \frac{4}{3}Tn_l\,,\nonumber\\
  \beta_1 &=& \frac{34}{3}C_A^2 - 4 C_F T n_l - \frac{20}{3} C_A T n_l\,.
\end{eqnarray}
Here $C_F=(N_c^2-1)/(2N_c)$ and $C_A=N_c$ are the quadratic Casimir operators of the
$SU(N_c)$ gauge group in the fundamental and adjoint representation,
respectively,  $n_l$ is the number of massless quark flavors, and $T=1/2$.

The perturbative expansions of $\Gamma_{\rm cusp}$ and $Z$ have the form
\begin{eqnarray}\label{z_minimal}
  \Gamma_{\rm cusp} &=& \sum_{i \ge 1} \Gamma_{\rm cusp}^{(i)}
  \left(\frac{{\alpha}_s}{ \pi}\right)^i \,,
  \nonumber\\
  Z &=& 1 + \sum_{ 1 \le j \le
    i } \frac{z_{i,j}}{\eps^j} \left(\frac{{\alpha}_s}{ \pi}\right)^i
  \,.
\end{eqnarray}
Solving Eq.~(\ref{rg_eq_ir}) to three loops, one finds
\begin{align}
  z_{1,1} =&  -\frac{1}{2} \Gamma^{(1)}_{\rm cusp}\,, \nonumber\\
  z_{2,2} =&  \frac{1}{16} \Gamma^{(1)}_{\rm cusp} ( \beta_0 + \Gamma^{(1)}_{\rm cusp})\,,  \nonumber\\
  z_{2,1} =&  -\frac{1}{4} \Gamma^{(2)}_{\rm cusp} \,,\nonumber\\
  z_{3,3} =& -\frac{1}{96} \Gamma^{(1)}_{\rm cusp} ( \beta_0 +
  \Gamma^{(1)}_{\rm cusp}) (\beta_0 + 2 \Gamma^{(1)}_{\rm cusp})\,, \nonumber\\
  z_{3,2} =& \frac{1}{96} ( \beta_1 \Gamma^{(1)}_{\rm cusp} + 4 \beta_0
  \Gamma^{(2)}_{\rm cusp} + 12 \Gamma^{(1)}_{\rm cusp} \Gamma^{(2)}_{\rm
    cusp}) \,, \nonumber\\
  z_{3,1} =& -\frac{1}{6} \Gamma^{(3)}_{\rm cusp} \,.
  \label{eq::F_poles}
\end{align}
The cusp anomalous dimension in QCD was computed to three loops
in~\cite{Polyakov:1980ca,Korchemsky:1987wg,Grozin:2014hna,Grozin:2015kna}. 

In this way, one can see explicitly the poles generated by the right-hand side
of Eq.~(\ref{ff_ir_ren}).  We have verified that this equation correctly
predicts all infrared poles in $F_1$ and $F_2$ to three loops.


\section{\label{sec::res}Results}


\subsection{Structure of results for form factors}

Before presenting explicit results we briefly discuss the general structure of
our analytic expressions.

All relevant master integrals were computed analytically in Ref.~\cite{HSS16}.
From this it is clear that the form factors are given in terms of iterated
integrals, with certain rational prefactors.  The required set of integration
kernels are
\begin{align}
  {\rm d}\log x \,,\qquad {\rm d}\log(1+x) \,,\qquad 
  {\rm d}\log(1-x)\,, \qquad {\rm d}\log(1-x+x^2)  \,.
\end{align} 
We sometimes refer to the arguments of the logarithms $x,1-x,1+x,1-x+x^2$ as
letters.

Up to two-loop order and for the three-loop fermionic contributions (i.e. the
$n_l^1$ and the $n_l^2$ terms) we observe only master integrals with letters
$x, 1-x$ and $1+x$. This means that all of them can be expressed in terms of
usual harmonic polylogarithms~\cite{Remiddi:1999ew,Maitre:2005uu}.

On the other hand, the non-fermionic three-loop part has the additional letter
$1-x+x^2$.  Introducing the complex roots of this polynomial, $r_{1,2} = (1\pm
i \sqrt{3})/2$, one can write ${\rm d}\log(1-x+x^2) = {\rm d}\log(x-r_1) +
{\rm d}\log(x-r_2)$.
In this way, all results can be written in terms of Goncharov polylogarithms.
See Ref.~\cite{HSS16} for more details.

At three-loop order we observe that $r_1=e^{i\pi/3}$ plays a special role for
the form factors $F_1$ and $F_2$ since the coefficients of the Goncharov
polylogarithms develop poles up to sixth order in $x-r_1$. We could show that
these poles are artificial by expanding the Goncharov polylogarithms around
$x=e^{i\pi/3}$. The analytic expressions for the 
finite result for $F_1$ and $F_2$ for $x=r_1$ are quite lengthy and can be
found (for $\mu^2=m^2$) in the ancillary file.


\subsection{\label{sub::ana}Analytical results}

We refrain from providing the results for the full three-loop form factors
since the analytic expressions are too lengthy.
All results which are discussed in this section can be downloaded
from~\verb|https://www.ttp.kit.edu/preprints/2016/ttp16-053/|.

It is instructive to consider the form factors $F_1$ and $F_2$ in various
kinematical regions which have already been mentioned in the Introduction.
They are discussed in the remaining part of this section.  In
Section~\ref{sec::num} they are numerically compared to the exact result.


\subsubsection{Low-energy: $s\ll m^2$ or $x\to1$}

We start with the limit $s\ll m^2$ which we obtain by expanding the Goncharov
polylogarithms in the master integrals for $x\to1$. The expansion has to be
carried out carefully since there are higher order poles in $1/(1-x)$ in the
prefactor.  In fact, we expand all master integrals up to order $(1-x)^9$ and
obtain $F_1$ and $F_2$ up to order $(1-x)^4$.  For the presentation in the
paper we write $x=e^{i\phi}$ and we
restrict ourselves to expansion terms up to order $\phi^2$, which for
$\mu^2=m^2$ are given by
\begin{eqnarray}
  F_1^{(1)} &=& \phi^2
\nc\Bigg[
-\frac{1}{3 \ep}-\frac{1}{4}
\Bigg]
  \,,\nonumber\\
  F_1^{(2)} &=& \phi^2\Bigg\{
 \nc\nl \Bigg[
-\frac{1}{9 \ep^2}+\frac{5}{27 \ep}+\frac{2 \pi ^2}{27}+\frac{283}{324}
\Bigg]
\nonumber\\&&\mbox{}
+ \nc^2\Bigg[
\frac{11}{18 \ep^2}+\frac{\frac{\pi ^2}{9}-\frac{47}{27}}{\ep}+\frac{4 \zeta
  (3)}{3}-\frac{5 \pi ^2}{54}
-\frac{5581}{1296}
\Bigg]
  \Bigg\}
  \,,\nonumber\\
  F_1^{(3)} &=& \phi^2\Bigg\{
   \nc\nl^2 \Bigg[
-\frac{4}{81 \ep^3}+\frac{20}{243 \ep^2}+\frac{4}{243 \ep}-\frac{56 \zeta
  (3)}{81}-\frac{58 \pi ^2}{243}
-\frac{7381}{4374}
\Bigg]
\nonumber\\&&\mbox{}
+ \nc^2\nl \Bigg[
\frac{44}{81 \ep^3}+\frac{\frac{4 \pi
    ^2}{81}-\frac{415}{243}}{\ep^2}+\frac{\frac{16 \zeta (3)}{27}+\frac{2051}{972}-\frac{40 \pi ^2}{243}}{\ep}+\frac{7 \zeta (3)}{27}-\frac{16 \pi^4}{81}+\frac{2380 \pi ^2}{
729}
\nonumber\\&&\mbox{}
+\frac{958687}{34992}
\Bigg]
+ \nc^3\Bigg[
-\frac{121}{81 \ep^3}+\frac{\frac{1340}{243}-\frac{22 \pi
    ^2}{81}}{\ep^2}+\frac{-\frac{10 \zeta (3)}{27}
-\frac{473}{54}+\frac{340 \pi ^2}{243}-\frac{2 \pi ^4}{27}}{\ep}
\nonumber\\&&\mbox{}
-5\zeta(5)
-\frac{46 \pi ^2 \zeta (3)}{9}+\frac{7127 \zeta (3)}{324}+\frac{70 \pi ^4}{81}-\frac{8977 \pi^2}{2916}-\frac{4961563}{69984}
\Bigg]
  \Bigg\}
  \,,\nonumber\\
  F_2^{(1)} &=& N_c \Bigg[ 1 + \frac{1}{6} \phi^2 \Bigg]
  \,,
  \nonumber\\
  F_1^{(2)} &=&
  \nc\nl \Bigg[
  -\frac{25}{9}
  \Bigg]
  + \nc^2\Bigg[
  \frac{355}{36}+\frac{2 \pi ^2}{3}
  \Bigg]
  \nonumber\\&&\mbox{}
  + \phi^2\Bigg\{
  + \nc\nl \Bigg[
  -\frac{31}{54}
  \Bigg]
  + \nc^2\Bigg[
  -\frac{1}{3 \ep}+\frac{23 \pi ^2}{90}-\frac{19}{27}
  \Bigg]
  \Bigg\}
  \,,\nonumber\\
  F_2^{(3)} &=&
  \nc\nl^2 \Bigg[
  \frac{634}{81}+\frac{8 \pi ^2}{27}
  \Bigg]
  + \nc^2\nl \Bigg[
  \frac{4 \zeta (3)}{3}-\frac{28451}{324}-6 \pi ^2
  \Bigg]
  \nonumber\\&&\mbox{}
  + \nc^3\Bigg[
  \frac{80 \zeta (3)}{3}+8 \pi ^2 \zeta (3)-20 \zeta
  (5)+\frac{104147}{648}+\frac{481 \pi ^2}{27}-\frac{2 
    \pi ^4}{3}
  \Bigg]
  \nonumber\\&&\mbox{}
  + \phi^2\Bigg\{    
  \nc\nl^2 \Bigg[
  \frac{467}{243}+\frac{4 \pi ^2}{81}
  \Bigg]
  + \nc^2\nl \Bigg[
  -\frac{1}{9 \ep^2}+\frac{2}{3 \ep}+\frac{4 \zeta (3)}{5}-\frac{7 \pi  ^2}{5}-\frac{190889}{19440}
  \Bigg]
  \nonumber\\&&\mbox{}
  + \nc^3\Bigg[
  \frac{11}{18 \ep^2}+\frac{-\frac{31}{12}-\frac{\pi ^2}{9}}{\ep}-6
  \zeta(5)+\frac{12 \pi ^2 \zeta (3)}{5}+\frac{407 \zeta (3)}{36}-\frac{23 \pi
    ^4}{90}+\frac{22849 \pi  ^2}{4320}
  \nonumber\\&&\mbox{}
  -\frac{3236461}{155520}
  \Bigg]
  \Bigg\}  
  \,.
\end{eqnarray}
Note that $F_2(x=1)$ is finite and agrees with Eqs.~(54) and~(55)
of Ref.~\cite{Grozin:2007fh} after adapting the large-$N_c$ limit.


\subsubsection{High-energy: $s\ll m^2$ or $x\to0$}

We expand all master integrals for $x\to0$ up to order $x^6$ which is
sufficient to obtain $F_1$ and $F_2$ up to order $x^4$. For illustration we
show the one-, two and three-loop results including the first power-suppressed
corrections of order $x^1$. It is convenient to write the $n$-loop component
of $F_i$ in the high-energy limit as follows
\begin{eqnarray}
  F_i^{(n)} &=& \sum_{k\ge0} f_{i,\rm lar}^{(n,k)} x^k
  \,.
  \label{eq::F_i_lar}
\end{eqnarray}
Our results for $F_1$ read (for $\mu^2=m^2$)
\begin{eqnarray}
  f_{1,\rm lar}^{(1,0)} &=&
 \nc\Bigg[
\left(-\frac{1}{\ep}-\frac{3}{2}\right)
 \logx-\frac{1}{\ep}-\frac{\logx^2}{2}+\frac{\pi ^2}{6}-2
\Bigg]
  \,,\nonumber\\
  f_{1,\rm lar}^{(2,0)} &=&
 \nc\nl \Bigg[
\left(-\frac{1}{3 \ep^2}+\frac{5}{9 \ep}+\frac{\pi ^2}{9}+\frac{209}{54}\right)
 \logx
\nonumber\\&&\mbox{}
-\frac{1}{3 \ep^2}+\frac{5}{9 \ep}+\frac{\logx^3}{9}+\frac{19 \logx^2}{18}-\frac{4 \zeta (3)}{3}-\frac{7 \pi ^2}{54}+\frac{106}{27}
\Bigg]
\nonumber\\&&\mbox{}
+ \nc^2\Bigg[
\left(\frac{1}{2 \ep^2}+\frac{2}{\ep}+\frac{29}{72}\right)
 \logx^2+\logx \left(\frac{17}{6 \ep^2}-\frac{2}{9 \ep}+5 \zeta (3)-\frac{11 \pi ^2}{18}-\frac{2795}{216}\right)
+\frac{7}{3 \ep^2}
\nonumber\\&&\mbox{}
+\left(\frac{1}{2 \ep}+\frac{19}{18}\right)
 \logx^3+\frac{-\zeta (3)-\frac{13}{18}}{\ep}+\frac{7 \logx^4}{24}+\frac{34 \zeta (3)}{3}-\frac{31 \pi ^4}{180}+\frac{337 \pi ^2}{216}-\frac{487}{27}
\Bigg]
  \,,\nonumber\\
  f_{1,\rm lar}^{(3,0)} &=&
 \nc\nl^2 \Bigg[
-\frac{4}{27 \ep^3}+\frac{20}{81 \ep^2}+\logx \left(-\frac{4}{27 \ep^3}+\frac{20}{81 \ep^2}+\frac{4}{81 \ep}-\frac{8 \zeta (3)}{27}-\frac{38 \pi ^2}{81}-\frac{4919}{729}\right)
\nonumber\\&&\mbox{}
+\frac{4}{81 \ep}-\frac{\logx^4}{27}-\frac{38 \logx^3}{81}+\left(-\frac{203}{81}-\frac{2 \pi ^2}{27}\right)
 \logx^2+\frac{32 \zeta (3)}{9}+\frac{29 \pi ^4}{405}-\frac{61 \pi ^2}{243}-\frac{3668}{729}
\Bigg]
\nonumber\\&&\mbox{}
+ \nc^2\nl \Bigg[
\frac{53}{27 \ep^3}+\left(\frac{1}{6 \ep^2}-\frac{41}{36 \ep}-\frac{5 \pi ^2}{24}-\frac{275}{81}\right)
 \logx^3+\frac{-\frac{4 \zeta (3)}{9}-\frac{316}{81}+\frac{\pi ^2}{54}}{\ep^2}
\nonumber\\&&\mbox{}
+\logx^2 \left(\frac{1}{3 \ep^3}+\frac{1}{9 \ep^2}+\frac{-\frac{481}{108}-\frac{5 \pi ^2}{36}}{\ep}-\frac{7 \zeta (3)}{9}-\frac{\pi ^2}{4}+\frac{4687}{648}\right)
\nonumber\\&&\mbox{}
+\logx \left(\frac{62}{27 \ep^3}+\frac{\frac{\pi
      ^2}{54}-\frac{713}{162}}{\ep^2}+\frac{\frac{22 \zeta
      (3)}{9}-\frac{961}{324}-\frac{71 \pi ^2}{324}}{\ep}-\frac{115 \zeta
    (3)}{9}+\frac{\pi ^4}{20}+\frac{8659 \pi ^2}{1944}
\right. \nonumber\\&&\mbox{} \left.
+\frac{56296}{729}\right)
+\left(-\frac{1}{18 \ep}-\frac{89}{72}\right)
 \logx^4+\frac{\frac{106 \zeta (3)}{27}-\frac{349}{324}-\frac{13 \pi
     ^2}{162}}{\ep}-\frac{\logx^5}{8}+\frac{164 \zeta (5)}{3}
\nonumber\\&&\mbox{}
-\frac{8 \pi ^2   \zeta (3)}{9}
-\frac{4684 \zeta (3)}{81}-\frac{293 \pi ^4}{1620}
-\frac{9979 \pi ^2}{1944}+\frac{500201}{5832}
\Bigg]
\nonumber\\&&\mbox{}
+ \nc^3\Bigg[
-\frac{175}{27 \ep^3}+\left(-\frac{1}{4 \ep^2}-\frac{10}{9 \ep}-\frac{\pi ^2}{9}+\frac{1141}{432}\right)
 \logx^4+\frac{\frac{31 \zeta (3)}{9}+\frac{1375}{162}-\frac{5 \pi
     ^2}{27}}{\ep^2}
\nonumber\\&&\mbox{}
+\logx^3 \left(-\frac{1}{6 \ep^3}-\frac{13}{6 \ep^2}+\frac{\frac{14}{9}-\frac{\pi ^2}{8}}{\ep}-\frac{49 \zeta (3)}{6}+\frac{37 \pi ^2}{72}+\frac{6373}{324}\right)
\nonumber\\&&\mbox{}
+\logx^2 \left(-\frac{7}{3 \ep^3}+\frac{-\frac{97}{36}-\frac{\pi
      ^2}{12}}{\ep^2}+\frac{-\frac{11 \zeta (3)}{2}+\frac{463}{27}+\frac{19
      \pi ^2}{72}}{\ep}-\frac{629 \zeta (3)}{36}-\frac{\pi ^4}{9}+\frac{593
    \pi ^2}{432}
\right. \nonumber\\&&\mbox{} \left.
+\frac{1535}{36}\right)
+\logx \left(-\frac{467}{54 \ep^3}+\frac{\zeta (3)+\frac{1645}{162}-\frac{29
      \pi ^2}{108}}{\ep^2}+\frac{-\frac{281 \zeta
      (3)}{18}+\frac{4369}{216}-\frac{161 \pi ^2}{648}+\frac{16 \pi
      ^4}{135}}{\ep}
\right. \nonumber\\&&\mbox{} \left.
+15 \zeta (5)-\frac{13 \pi ^2 \zeta (3)}{36}+\frac{4297 \zeta (3)}{54}-\frac{71 \pi ^4}{360}-\frac{31609 \pi ^2}{1944}-\frac{669127}{5832}\right)
\nonumber\\&&\mbox{}
+\left(-\frac{5}{24 \ep}-\frac{3}{8}\right)
 \logx^5+\frac{-\frac{550 \zeta (3)}{27}-\frac{7 \pi ^2 \zeta (3)}{18}+6 \zeta
   (5)+\frac{637}{54}-\frac{161 \pi ^2}{648}+\frac{16 \pi
     ^4}{135}}{\ep}-\frac{\logx^6}{8} 
\nonumber\\&&\mbox{}
-\frac{875 \zeta (5)}{3}-\frac{16 \zeta (3)^2}{3}+\frac{113 \pi ^2 \zeta
  (3)}{18}+\frac{33197 \zeta (3)}{162}+\frac{2039 \pi ^6}{17010}-\frac{1727
  \pi ^4}{1080}
\nonumber\\&&\mbox{}
+\frac{23773 \pi ^2}{1296}-\frac{554267}{2916}
\Bigg] 
  \,,\nonumber\\
  f_{1,\rm lar}^{(1,1)} &=&
 \nc\Bigg[
\logx-2
\Bigg]
  \,,\nonumber\\
  f_{1,\rm lar}^{(2,1)} &=&
 \nc\nl \Bigg[
-\frac{\logx^2}{3}-\frac{37 \logx}{9}+\frac{\pi ^2}{9}+\frac{50}{9}
\Bigg]
+ \nc^2\Bigg[
\left(-\frac{1}{\ep}+\pi ^2-\frac{161}{12}\right)
 \logx^2
\nonumber\\&&\mbox{}
+\logx \left(\frac{1}{\ep}-48 \zeta (3)+\frac{31 \pi ^2}{6}+\frac{799}{36}\right)
+\frac{2}{\ep}-\logx^3-122 \zeta (3)+\frac{17 \pi ^4}{30}
\nonumber\\&&\mbox{}
+\frac{449 \pi ^2}{36}-\frac{1003}{18}
\Bigg]
  \,,\nonumber\\
  f_{1,\rm lar}^{(3,1)} &=&
 \nc\nl^2 \Bigg[
\frac{4 \logx^3}{27}+\frac{74 \logx^2}{27}+\left(\frac{1090}{81}+\frac{4 \pi ^2}{27}\right)
 \logx-\frac{16 \zeta (3)}{9}-\frac{122 \pi ^2}{81}-\frac{1412}{81}
\Bigg]
\nonumber\\&&\mbox{}
+ \nc^2\nl \Bigg[
\logx^2 \left(-\frac{1}{3 \ep^2}+\frac{23}{6 \ep}+20 \zeta (3)-\frac{203 \pi ^2}{36}+\frac{1307}{27}\right)
\nonumber\\&&\mbox{}
+\logx \left(\frac{1}{3 \ep^2}+\frac{-\frac{4}{3}-\frac{\pi ^2}{18}}{\ep}+\frac{448 \zeta (3)}{3}+\frac{34 \pi ^4}{45}-\frac{3823 \pi ^2}{108}-\frac{31109}{324}\right)
\nonumber\\&&\mbox{}
+\frac{2}{3 \ep^2}+\left(\frac{1}{6 \ep}-\frac{2 \pi ^2}{3}+\frac{779}{54}\right)
 \logx^3+\frac{-5-\frac{\pi ^2}{18}}{\ep}+\frac{5 \logx^4}{9}-208 \zeta (5) 
\nonumber\\&&\mbox{}
+464 \zeta (3)
+\frac{101 \pi ^4}{270}-\frac{2972 \pi ^2}{81}+\frac{44921}{162}
\Bigg]
\nonumber\\&&\mbox{}
+ \nc^3\Bigg[
\logx^3 \left(\frac{1}{2 \ep^2}+\frac{\frac{40}{3}-\pi ^2}{\ep}+40 \zeta (3)-\frac{197 \pi ^2}{72}-\frac{3095}{108}\right)
\nonumber\\&&\mbox{}
+\logx^2 \left(\frac{11}{6 \ep^2}+\frac{48 \zeta (3)-\frac{41}{6}-\frac{71 \pi ^2}{12}}{\ep}-130 \zeta (3)+\frac{17 \pi ^4}{10}+\frac{11 \pi ^2}{9}-\frac{13525}{108}\right)
\nonumber\\&&\mbox{}
+\logx \left(-\frac{10}{3 \ep^2}+\frac{169 \zeta (3)+\frac{457}{12}-\frac{323
      \pi ^2}{18}-\frac{17 \pi ^4}{30}}{\ep}+96 \zeta (5)+2 \pi ^2 \zeta
  (3)-\frac{3391 \zeta (3)}{3}
\right. \nonumber\\&&\mbox{} \left.
+\frac{401 \pi ^4}{72}+\frac{8221 \pi ^2}{72}-\frac{1451}{81}\right)
-\frac{14}{3 \ep^2}+\left(\frac{3}{4 \ep}-\frac{3 \pi ^2}{2}+\frac{1319}{72}\right)
 \logx^4
\nonumber\\&&\mbox{}
+\frac{124 \zeta (3)+\frac{109}{2}-\frac{469 \pi ^2}{36}-\frac{17 \pi
    ^4}{30}}{\ep}+\frac{5 \logx^5}{8}+2154 \zeta (5)+18 \zeta (3)^2-\frac{139
  \pi ^2 \zeta (3)}{3}
\nonumber\\&&\mbox{}
-\frac{24647 \zeta (3)}{9}-\frac{799 \pi ^6}{1260}+\frac{5291 \pi ^4}{2160}+\frac{122195 \pi ^2}{648}-\frac{149093}{324}
\Bigg]
  \,,
\end{eqnarray}
with $\logx=\log(x)$.
For $F_2$ we get the following expansion coefficients 
\begin{eqnarray}
  f_{2,\rm lar}^{(1,0)} &=& 0
  \,,\nonumber\\
  f_{2,\rm lar}^{(2,0)} &=& 0
  \,,\nonumber\\
  f_{2,\rm lar}^{(3,0)} &=& 0
  \,,\nonumber\\
  f_{2,\rm lar}^{(1,1)} &=&
  -2 \logx \nc
  \,,\nonumber\\
  f_{2,\rm lar}^{(2,1)} &=&
 \nc\nl \Bigg[
\frac{2 \logx^2}{3}+\frac{50 \logx}{9}-\frac{2 \pi ^2}{9}
\Bigg]
+ \nc^2\Bigg[
\left(\frac{2}{\ep}+\frac{53}{6}\right)
 \logx^2+\left(\frac{2}{\ep}-2 \pi ^2-\frac{67}{18}\right)
 \logx+2 \logx^3
\nonumber\\&&\mbox{}
+44 \zeta (3)-\frac{77 \pi ^2}{18}+6
\Bigg]
  \,,\nonumber\\
  f_{2,\rm lar}^{(3,1)} &=&
 \nc\nl^2 \Bigg[
-\frac{8 \logx^3}{27}-\frac{100 \logx^2}{27}+\left(-\frac{1268}{81}-\frac{8 \pi ^2}{27}\right)
 \logx+\frac{32 \zeta (3)}{9}+\frac{100 \pi ^2}{81}
\Bigg]
\nonumber\\&&\mbox{}
+ \nc^2\nl \Bigg[
\left(\frac{2}{3 \ep^2}-\frac{13}{3 \ep}+\frac{13 \pi ^2}{18}-\frac{787}{27}\right)
 \logx^2
\nonumber\\&&\mbox{}
+\logx \left(\frac{2}{3 \ep^2}+\frac{\frac{\pi ^2}{9}-4}{\ep}-\frac{76 \zeta (3)}{3}+\frac{109 \pi ^2}{6}+\frac{12773}{162}\right)
\nonumber\\&&\mbox{}
+\left(-\frac{1}{3 \ep}-\frac{296}{27}\right)
 \logx^3+\frac{\pi ^2}{9 \ep}-\frac{10 \logx^4}{9}-\frac{2060 \zeta (3)}{9}-\frac{8 \pi ^4}{135}+\frac{1000 \pi ^2}{81}-\frac{122}{3}
\Bigg]
\nonumber\\&&\mbox{}
+ \nc^3\Bigg[
\left(-\frac{1}{\ep^2}-\frac{23}{3 \ep}+\frac{19 \pi ^2}{12}+\frac{277}{27}\right)
 \logx^3
\nonumber\\&&\mbox{}
+\logx^2 \left(-\frac{17}{3 \ep^2}+\frac{\frac{3 \pi ^2}{2}-\frac{17}{3}}{\ep}-13 \zeta (3)+\frac{133 \pi ^2}{36}+\frac{7201}{54}\right)
\nonumber\\&&\mbox{}
+\logx \left(-\frac{14}{3 \ep^2}+\frac{-42 \zeta (3)-\frac{15}{2}+\frac{56 \pi ^2}{9}}{\ep}+\frac{754 \zeta (3)}{3}-\frac{41 \pi ^4}{30}-\frac{2050 \pi ^2}{27}+\frac{2255}{162}\right)
\nonumber\\&&\mbox{}
+\left(-\frac{3}{2 \ep}-\frac{227}{36}\right)
 \logx^4+\frac{-44 \zeta (3)-6+\frac{85 \pi ^2}{18}}{\ep}-\frac{5
   \logx^5}{4}-732 \zeta (5)+\frac{77 \pi ^2 \zeta (3)}{3}
\nonumber\\&&\mbox{}
+\frac{4697 \zeta   (3)}{3}
+\frac{329 \pi ^4}{540}-\frac{39005 \pi ^2}{324}+\frac{548}{3}
\Bigg]
  \,.
\end{eqnarray}

Note that the coefficients $f_{i,\rm lar}^{(n,k)}$ contain logarithmic terms
in $x$ which leads to a divergent behaviour of $F_i^{(n)}$ for $x\to0$. For
this reason we subtract $f_{i,\rm lar}^{(n,0)}$ when comparing with the exact
result (cf Section~\ref{sec::num}).
In Ref.~\cite{Gluza:2009yy} some of the pole parts for $F_1$ of the leading
three-loop coefficient $f_{1,\rm lar}^{(3,0)}$ have been predicted using
evolution equations. However, the $\epsilon^0$ term, the sub-leading terms
of order $x^n$ with $n\ge1$, and the results for $f_{2,\rm lar}^{(3,n)}$ 
are new (see also Subsection~\ref{sub::checks}).
Finally, we want to remark that higher order $\epsilon$ terms for the
one- and two-loop coefficients can be found in the ancillary file.


\subsubsection{Threshold: $s\to4m^2$ or $x\to-1$}

To obtain the threshold limit we expand the master integrals up to order
$(1+x)^6$. After inserting the expanded results into the expressions for the
form factors it is convenient to use
\begin{eqnarray}
  x &=& \frac{2\beta}{1+\beta} - 1
  \,,
\end{eqnarray}
where 
\begin{eqnarray}
  \beta &=& \sqrt{1 - \frac{4 m^2}{s}}
  \,,
\end{eqnarray}
is the velocity of the produced heavy quarks. Note that
the ultravioletly renormalized form factors develop poles up to order
$1/\beta^{n}$ where $n=1,2,3$ is the number of loops.  On the other hand, the
bare form factors have poles up to $1/\beta^{2n}$ (cf. Ref.~\cite{Gluza:2009yy}
where bare two-loop results are presented).  Since the resulting expressions
are quite large we refrain from displaying them in the paper but refer to the
ancillary file which comes together with this paper.
It is, however, instructive to look into the
cross section $\sigma(e^+e^-\to Q \bar{Q})$, where $Q$ is a heavy quark.
Close to threshold it is determined by the
virtual correction, i.e., the form factors $F_1$ and $F_2$, since
the contributions from real radiation are suppressed by a relative factor 
$\beta^3$. In fact, we can write
\begin{eqnarray}
  \sigma(e^+e^-\to Q \bar{Q}) &=& \sigma_0 \beta
  \left[ |F_1 + F_2|^2 
    + \frac{ |(1-\beta^2) F_1 + F_2 |^2}{2(1-\beta^2)}
  \right] 
  \nonumber\\
  &=& \sigma_0 \frac{3\beta}{2}\left[
    1-\frac{\beta^2}{3}
    + \frac{\alpha_s}{4\pi} \Delta^{(1)} 
    + \left(\frac{\alpha_s}{4\pi}\right)^2 \Delta^{(2)}
    + \left(\frac{\alpha_s}{4\pi}\right)^3 \Delta^{(3)}
    + \ldots
  \right]
  \,.
  \label{eq::sigee}
\end{eqnarray}
where $\sigma_0 = 4\pi \alpha^2 Q_Q^2 /(3s)$.  Our calculation of $F_1$ and
$F_2$ determines the first three terms for each $\Delta^{(n)}$ in the
expansion for $\beta\to 0$.  Note that individually $F_1$ and $F_2$ still
contain poles in $\epsilon$, however, the combination given in
Eq.~(\ref{eq::sigee}) is finite. For the one-, two- and three-loop corrections
we have
\begin{eqnarray}
  \Delta^{(1)} &=& 
  \nc\Bigg[
  \frac{\pi ^2}{\beta} -8 + \beta \frac{2 \pi ^2}{3}
  \Bigg]
  +\ldots
  \,,\nonumber\\
  \Delta^{(2)} &=&
  \nc\nl \Bigg[
  \frac{1}{\beta} \Bigg(\frac{4}{3} \pi ^2 \log (2 \beta )-\frac{10 \pi ^2}{9}\Bigg)+ \frac{44}{9}
  \Bigg]
  \nonumber\\&&\mbox{}
  + \nc^2\Bigg[
  \frac{\pi ^4}{3\beta^2} + \frac{1}{\beta}
  \Bigg(-\frac{22}{3} \pi ^2 \log (2 \beta )-\frac{41 \pi ^2}{9}\Bigg)
  \nonumber\\&&\mbox{}
  -\frac{32}{3} \pi ^2 \log (2 \beta )-56 \zeta (3)+\frac{5 \pi ^4}{9}+\frac{109 \pi ^2}{9}+\frac{49}{9}-\frac{16}{3} \pi ^2 \log (2)
  \Bigg]
  +\ldots\,,
  \nonumber\\
  \Delta^{(3)} &=&
  \nc\nl^2 
  \frac{1}{\beta} \Bigg(\frac{16}{9} \pi ^2 \log ^2(2 \beta )-\frac{80}{27} \pi ^2 \log (2 \beta )+\frac{8 \pi ^4}{27}+\frac{100 \pi ^2}{81}\Bigg)
  \nonumber\\&&\mbox{}
  + \nc^2\nl \Bigg[
  \frac{1}{\beta^2} \Bigg(\frac{8}{9} \pi ^4 \log (2 \beta )+\frac{16 \pi ^2
    \zeta (3)}{3}-\frac{20 \pi ^4}{27}\Bigg)
  \nonumber\\&&\mbox{}
  + \frac{1}{\beta} \Bigg(-\frac{176}{9} \pi ^2 \log ^2(2 \beta )+\frac{634}{27} \pi ^2 \log (2 \beta )-\frac{16 \pi ^2 \zeta (3)}{3}-\frac{88 \pi ^4}{27}-\frac{617 \pi ^2}{324}\Bigg)
  \Bigg]
  \nonumber\\&&\mbox{}
  + \nc^3\Bigg[
  \frac{1}{\beta^2} \Bigg(-\frac{44}{9} \pi ^4 \log (2 \beta )-\frac{88 \pi ^2
    \zeta (3)}{3}-\frac{10 \pi ^4}{27}\Bigg)
  \nonumber\\&&\mbox{}
  + \frac{1}{\beta} \Bigg(\frac{484}{9} \pi ^2 \log ^2(2 \beta )-\frac{32}{3}
  \pi ^4 \log (2 \beta )-\frac{392}{27} \pi ^2 \log (2 \beta )-\frac{146 \pi ^2
    \zeta (3)}{3}-\frac{\pi ^6}{4}
  \nonumber\\&&\mbox{}
  +\frac{677 \pi ^4}{27}+\frac{761 \pi ^2}{162}-\frac{16}{3} \pi ^4 \log (2)\Bigg)
  \Bigg]
  +\ldots  \,,
  \label{eq::Delta}
\end{eqnarray}
where the ellipses refer to higher order terms in $\beta$.  The one- and
two-loop expressions agree with the large-$N_c$ limit of
Refs.~\cite{Kallen:1955fb,Czarnecki:1997vz,Beneke:1997jm} and the three-loop
terms agree with
Ref.~\cite{Pineda:2006ri,Hoang:2008qy,Kiyo:2009gb}.\footnote{We thank Andreas
  Maier for providing the result for $\Pi^{(3),v}(z)$ in Eq.~(A.6) of
  Ref.~\cite{Kiyo:2009gb} and the corresponding two-loop expression in terms
  of Casimir invariants.}  At $n$-loop order the leading term of
$\Delta^{(n)}$ behaves as $(\alpha_s/\beta)^n$ which is determined by the
Sommerfeld factor~\cite{textbook} $S = z/(1-e^{-z})$ with $z =
C_F\alpha_s\pi/\beta$. It is interesting to note that the series expansion of
$S$ has no term of order $\alpha_s^3$ and thus $\Delta^{(3)}$ starts at order
$1/\beta^2$ which is confirmed by our explicit calculation.

In the context of effective theories an important quantity derived from the
form factor $F_1$ is the matching coefficient between QCD and non-relativistic
QCD of the vector current. It is obtained by considering the on-shell
photon-quark vertex for $q^2=4m^2$. Due to the singularities in $1/\beta$ (see
above) it is not possible to obtain the matching coefficient from the general
result for $F_1$. Rather a dedicated calculation is necessary which has been
performed in~\cite{Marquard:2014pea} to three-loop order using semi-analytical
methods.  The planar master integrals of~\cite{Marquard:2014pea} have been
computed in~\cite{HSS16} as by-product of the calculation of all master
integrals used in this calculation.



\subsection{\label{sec::num}Numerical results}

\begin{figure}[t] 
  \begin{center}
    \includegraphics[width=.3\textwidth]{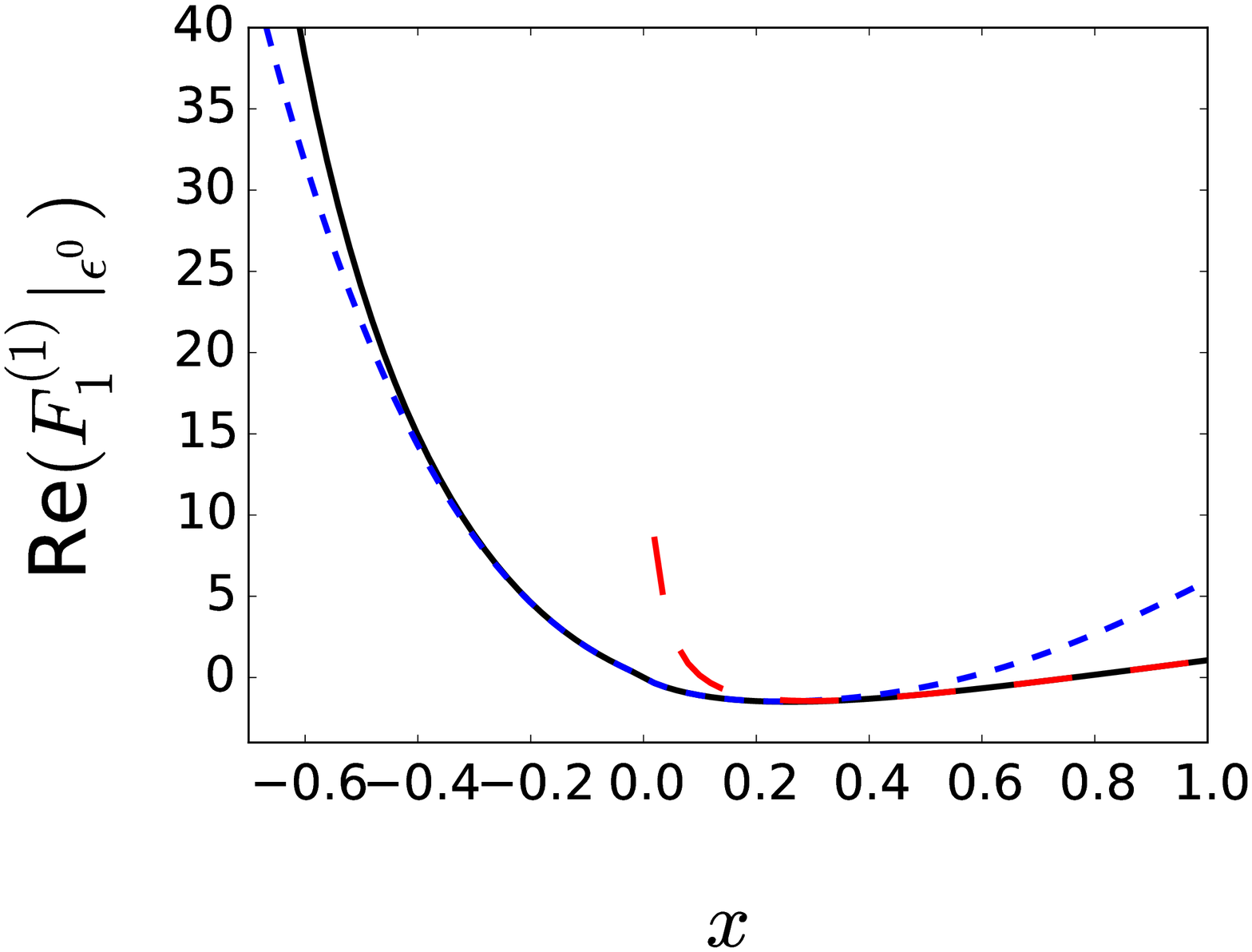}
    \includegraphics[width=.3\textwidth]{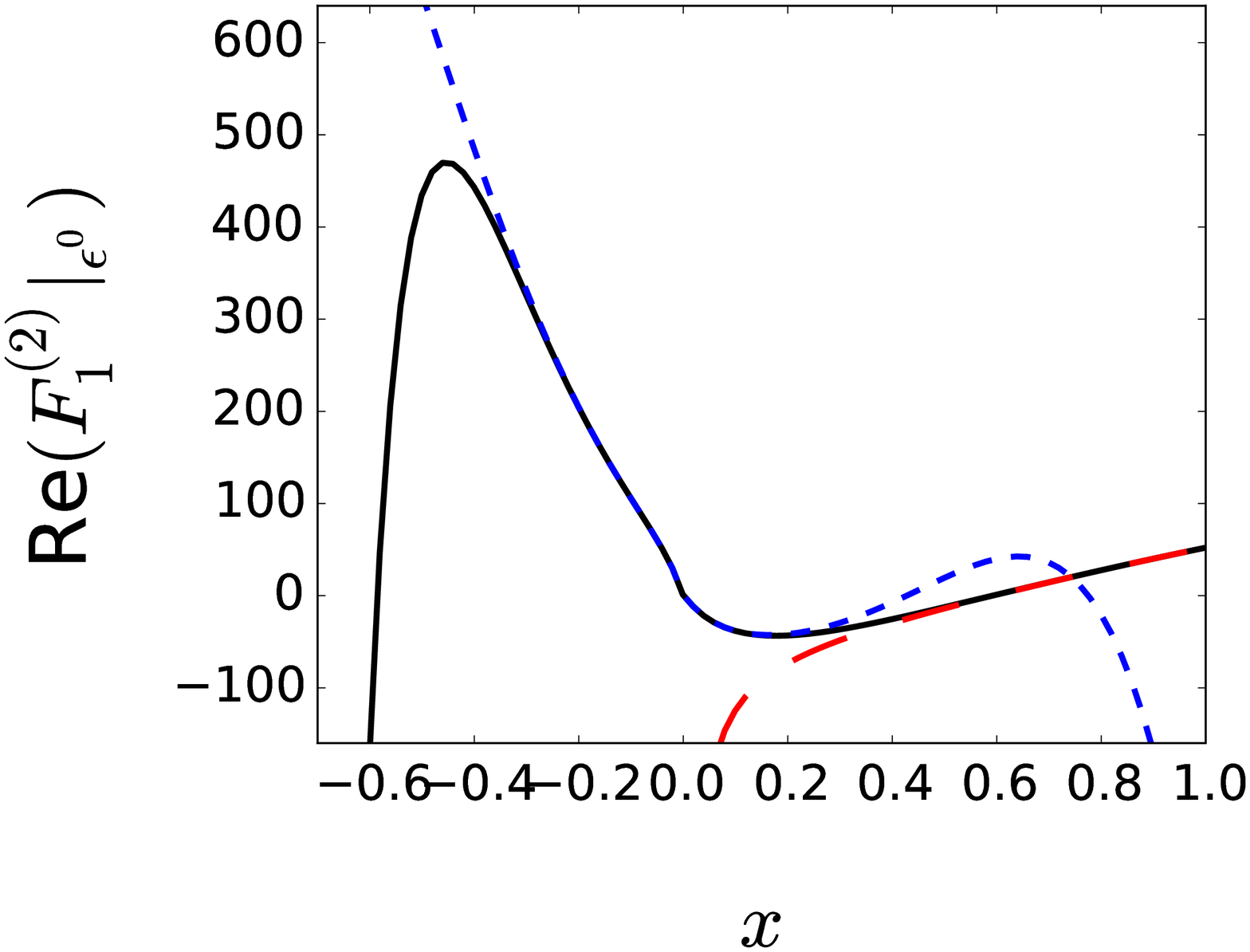}
    \includegraphics[width=.3\textwidth]{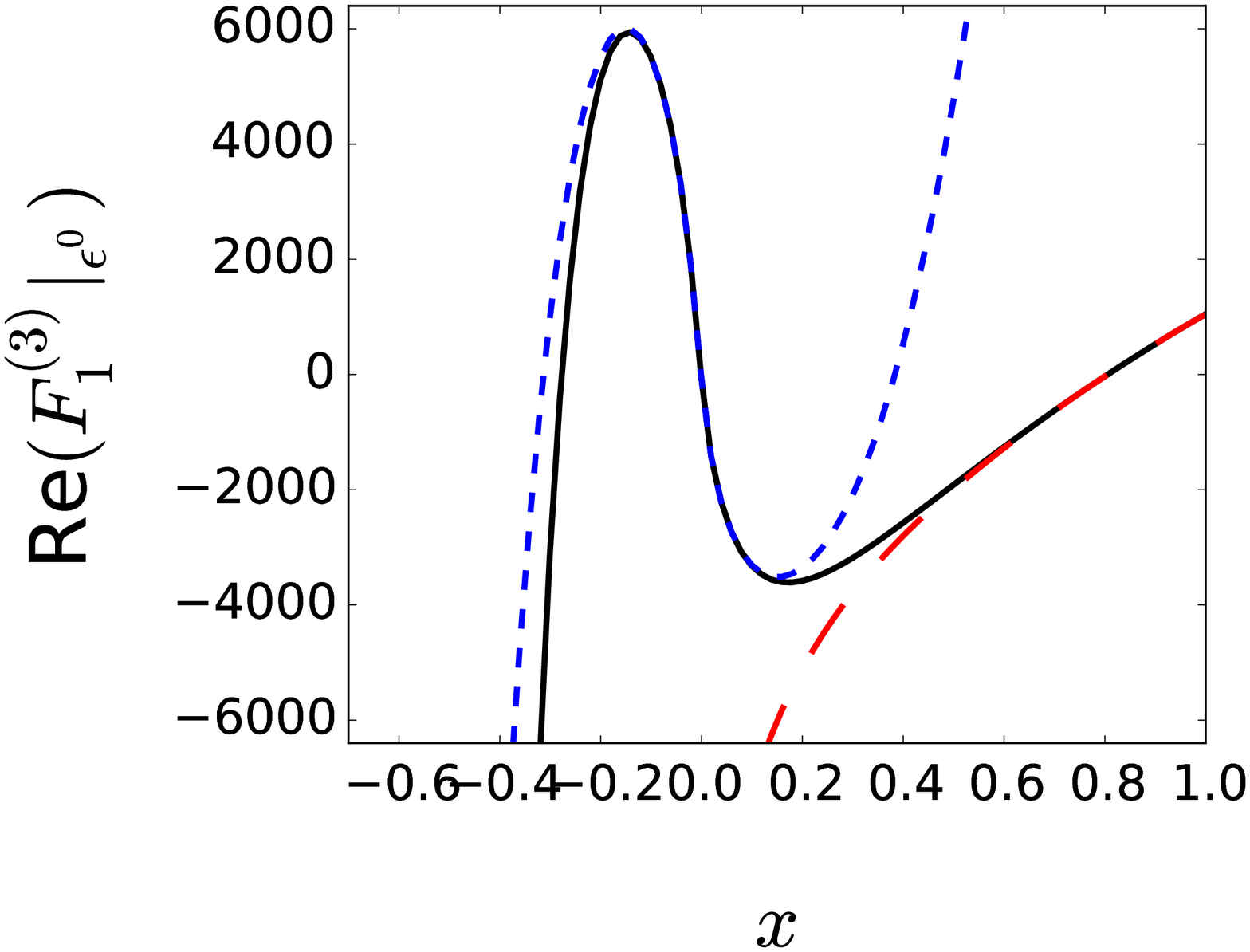}
    \\
    \includegraphics[width=.3\textwidth]{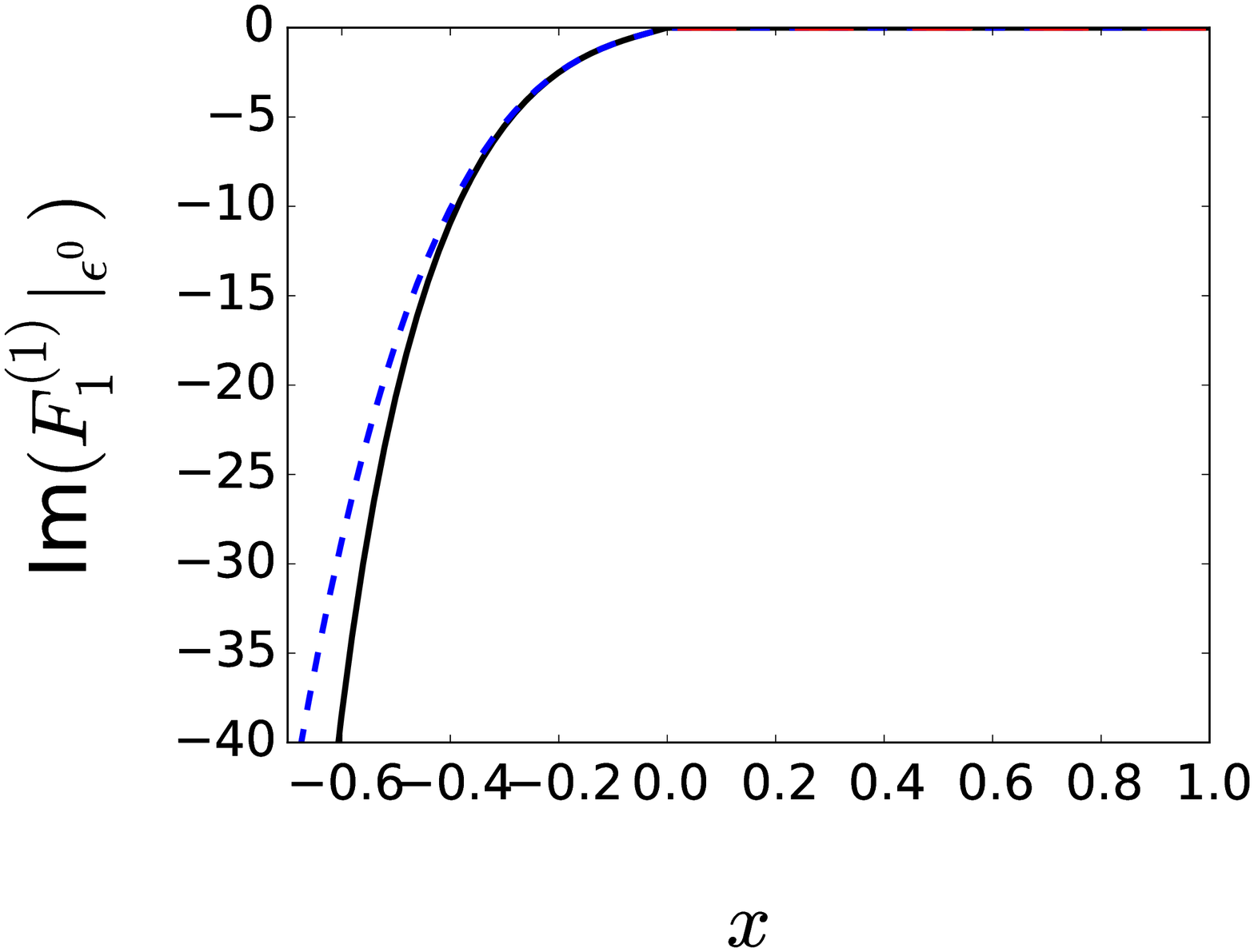}
    \includegraphics[width=.3\textwidth]{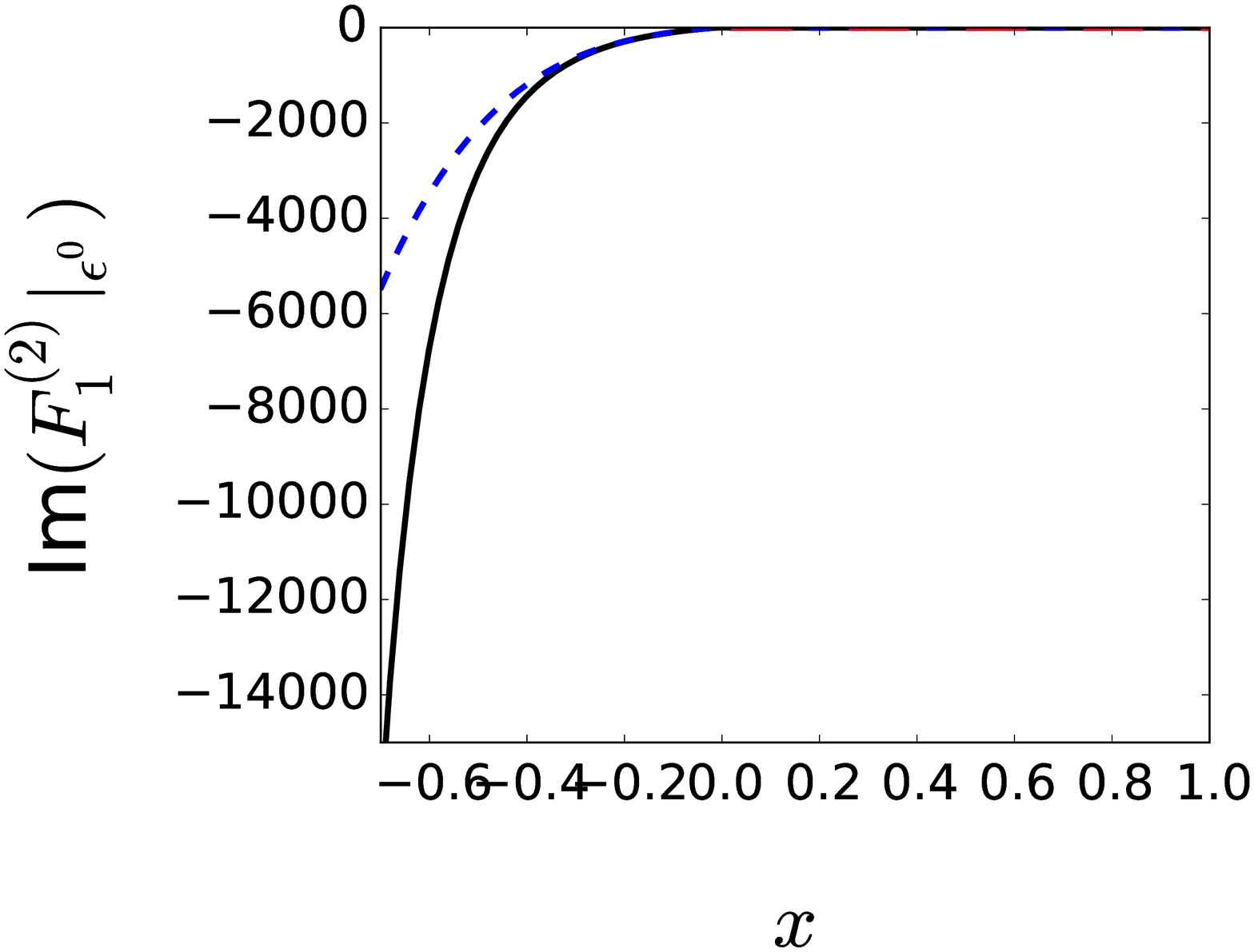}
    \includegraphics[width=.3\textwidth]{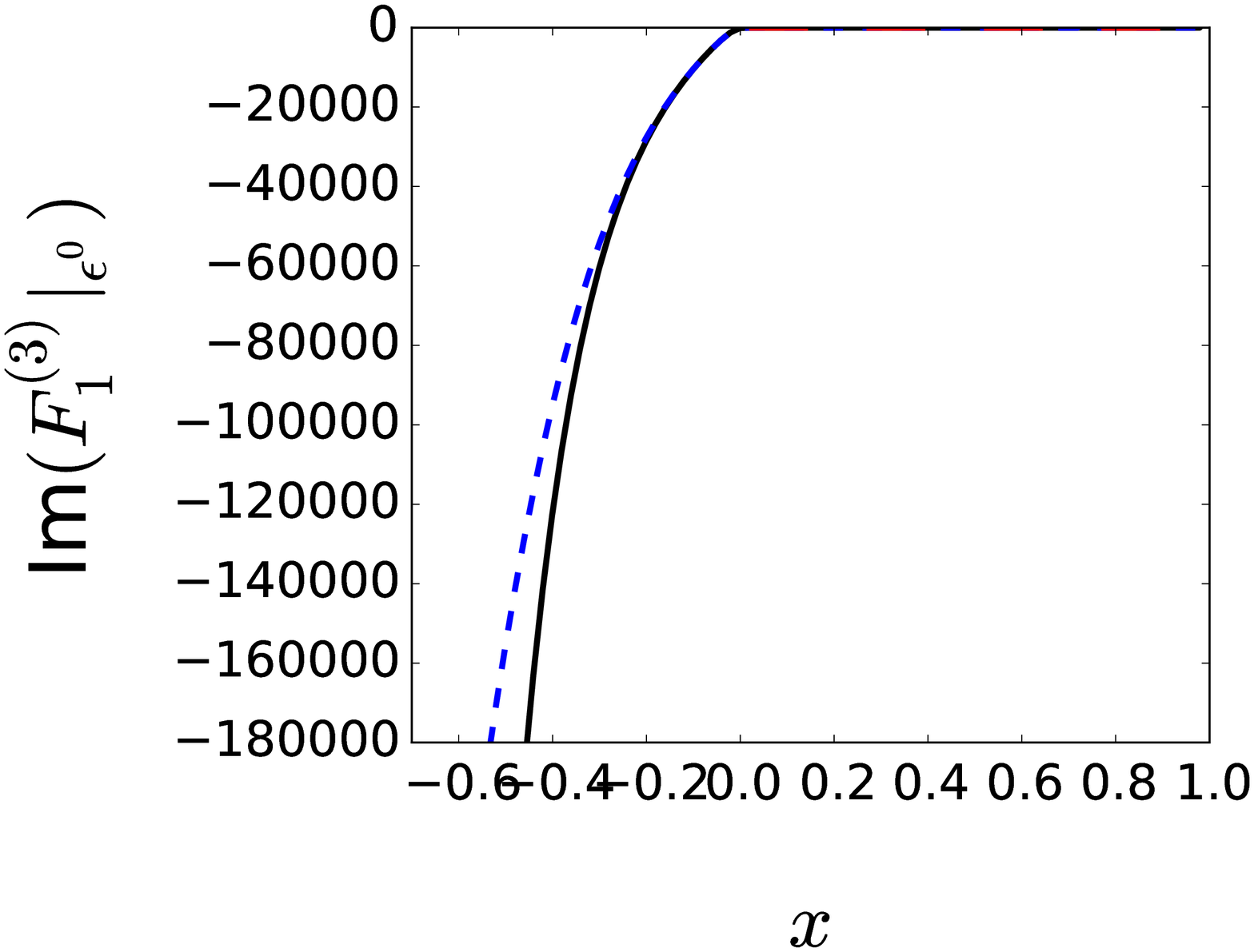}
    \caption{\label{fig::F1_x}Real and imaginary parts of $\epsilon^0$ one-,
      two- and three-loop contribution of $F_1$ as a function of $x$. The
      leading high-energy term (i.e. $f_{1,\rm lar}^{(n,0)}$ from
      Eq.~(\ref{eq::F_i_lar})) is subtracted so that $F_1$ is zero for $x=0$.
      The solid (black) lines show the exact result and the short-dashed
      (blue) lines represent the high-energy approximations including terms up
      to order $x^4$.  The long-dashed (red) curves contain low-energy
      expansion terms up to order $(1-x)^4$.  The number of light fermions is
      set to zero ($n_l=0$).}
  \end{center}
\end{figure}

\begin{figure}[t] 
  \begin{center}
    \includegraphics[width=.3\textwidth]{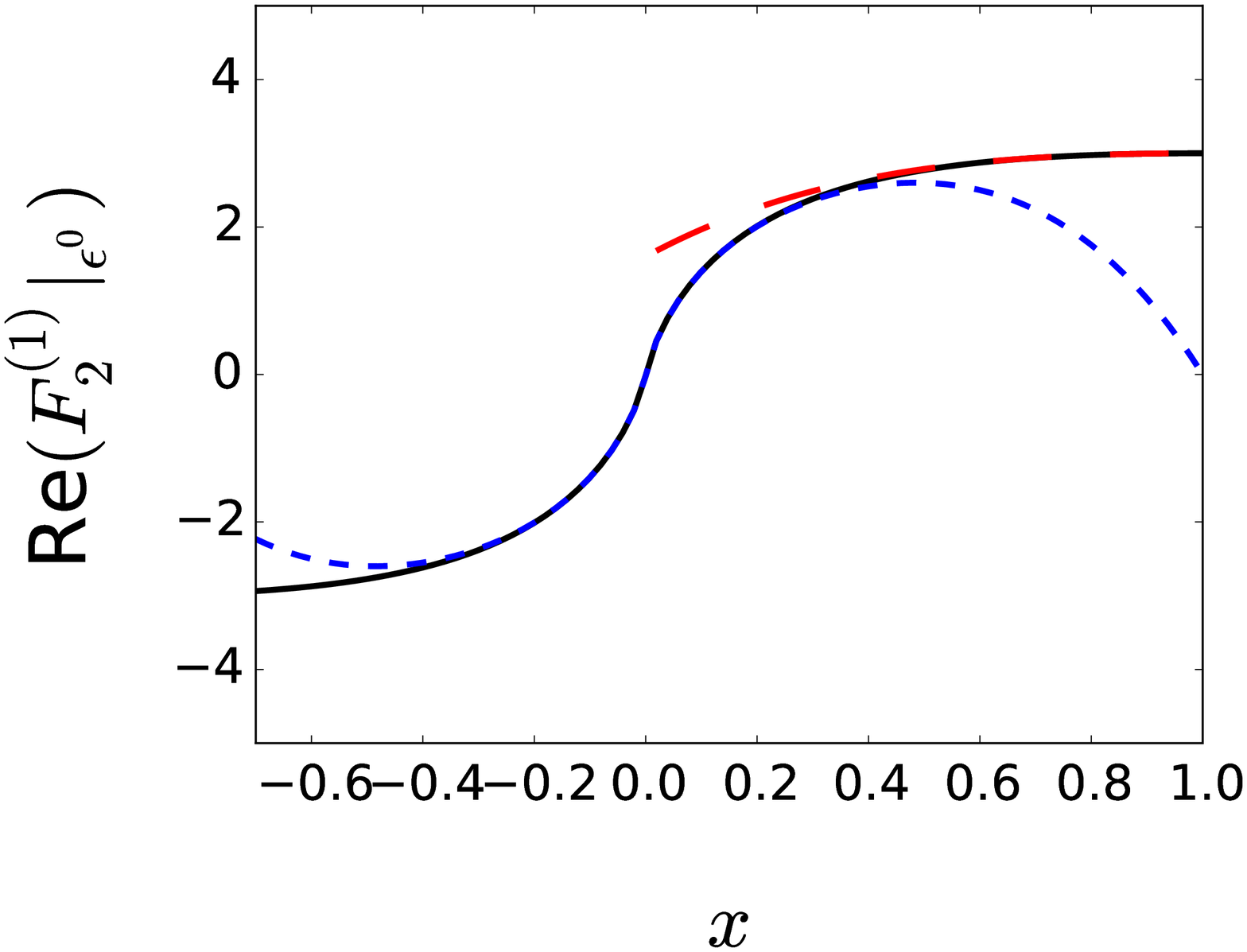}
    \includegraphics[width=.3\textwidth]{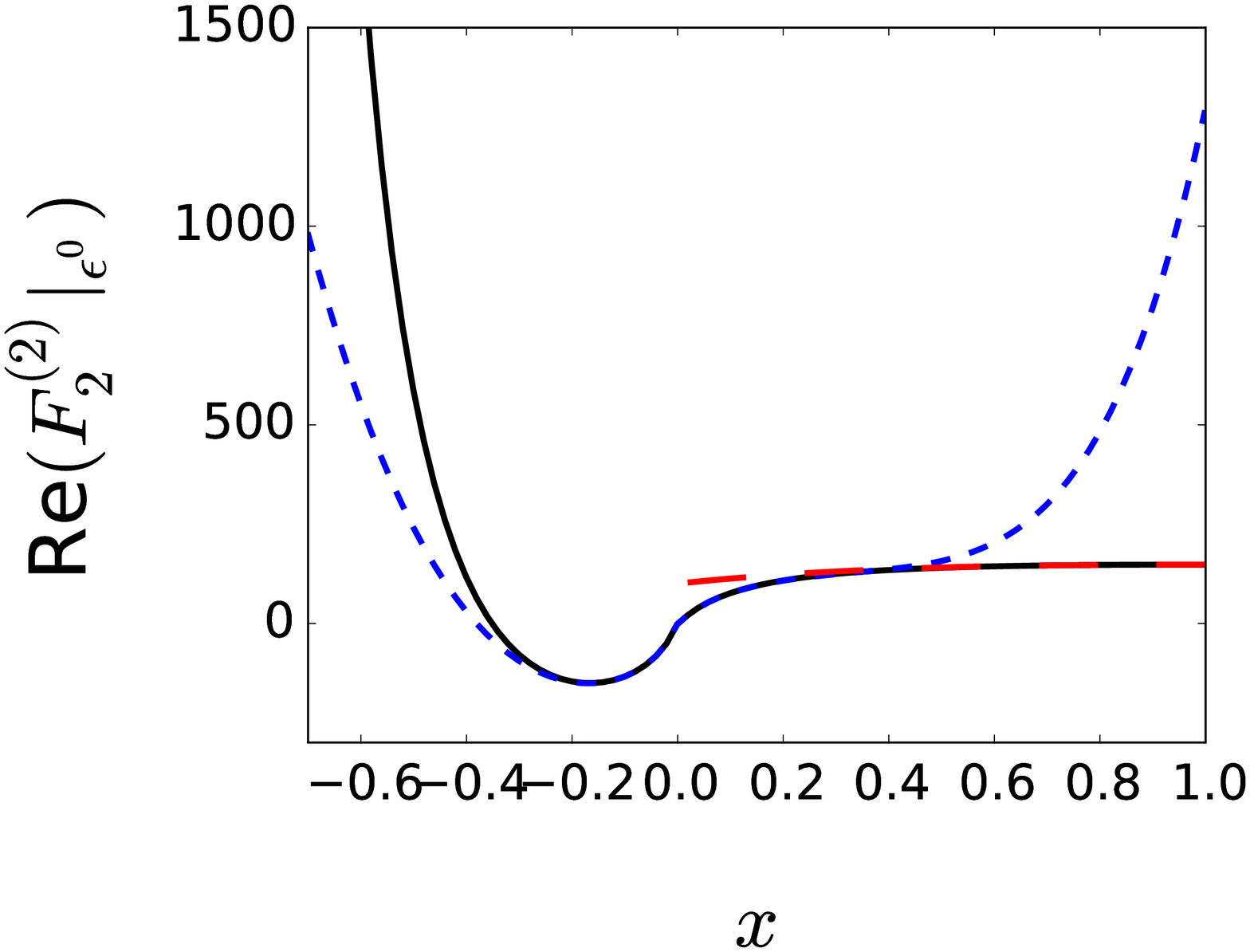}
    \includegraphics[width=.3\textwidth]{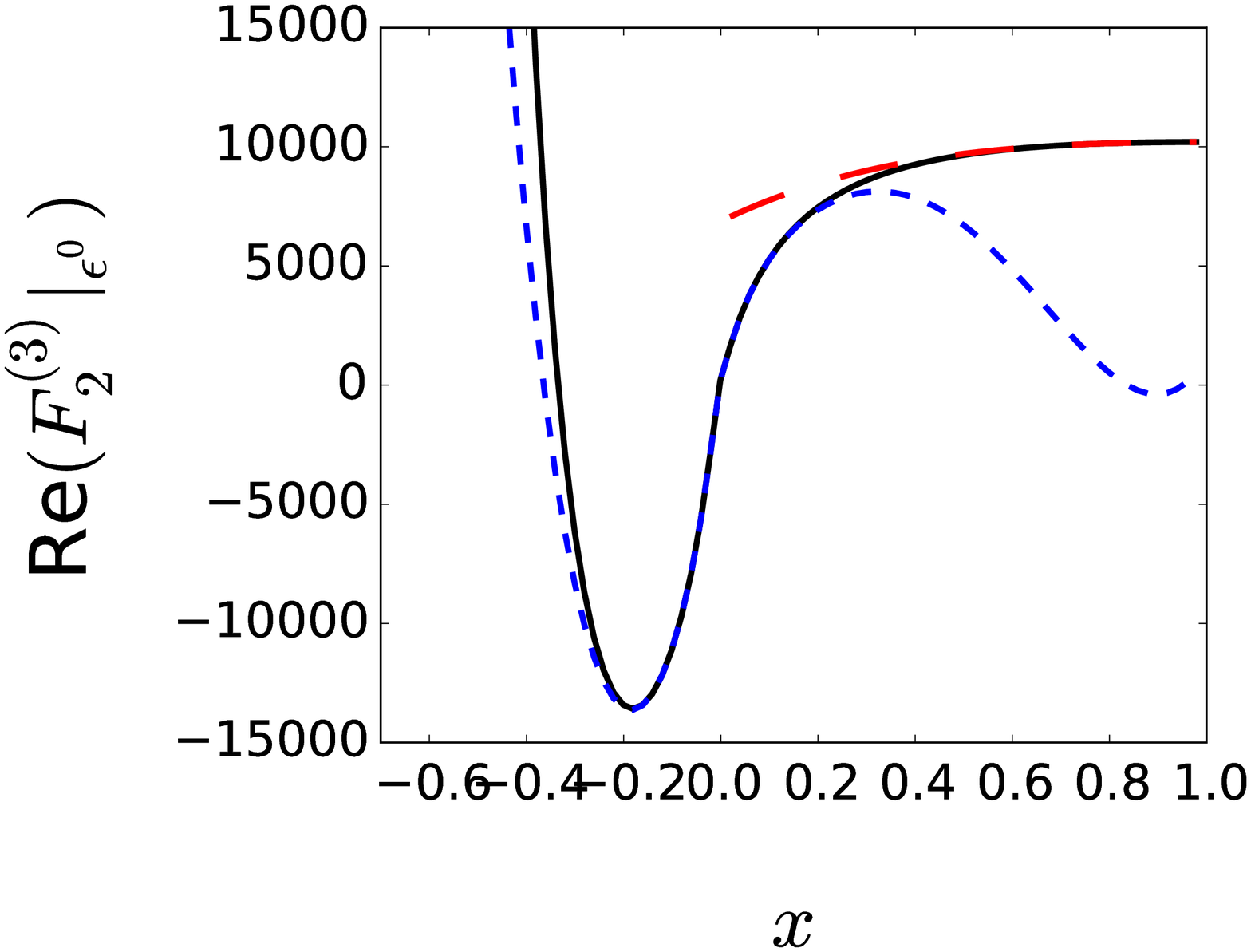}
    \\
    \includegraphics[width=.3\textwidth]{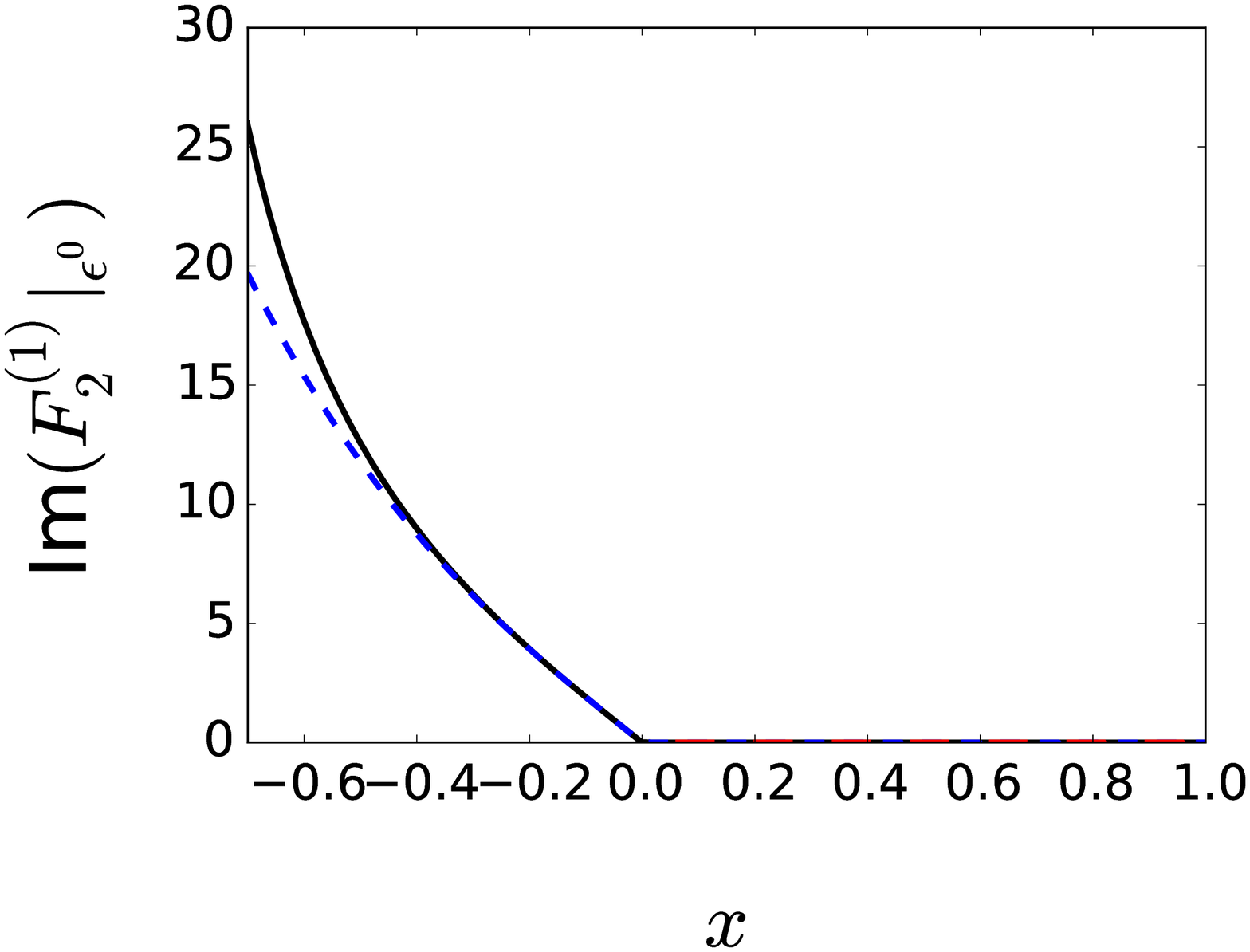}
    \includegraphics[width=.3\textwidth]{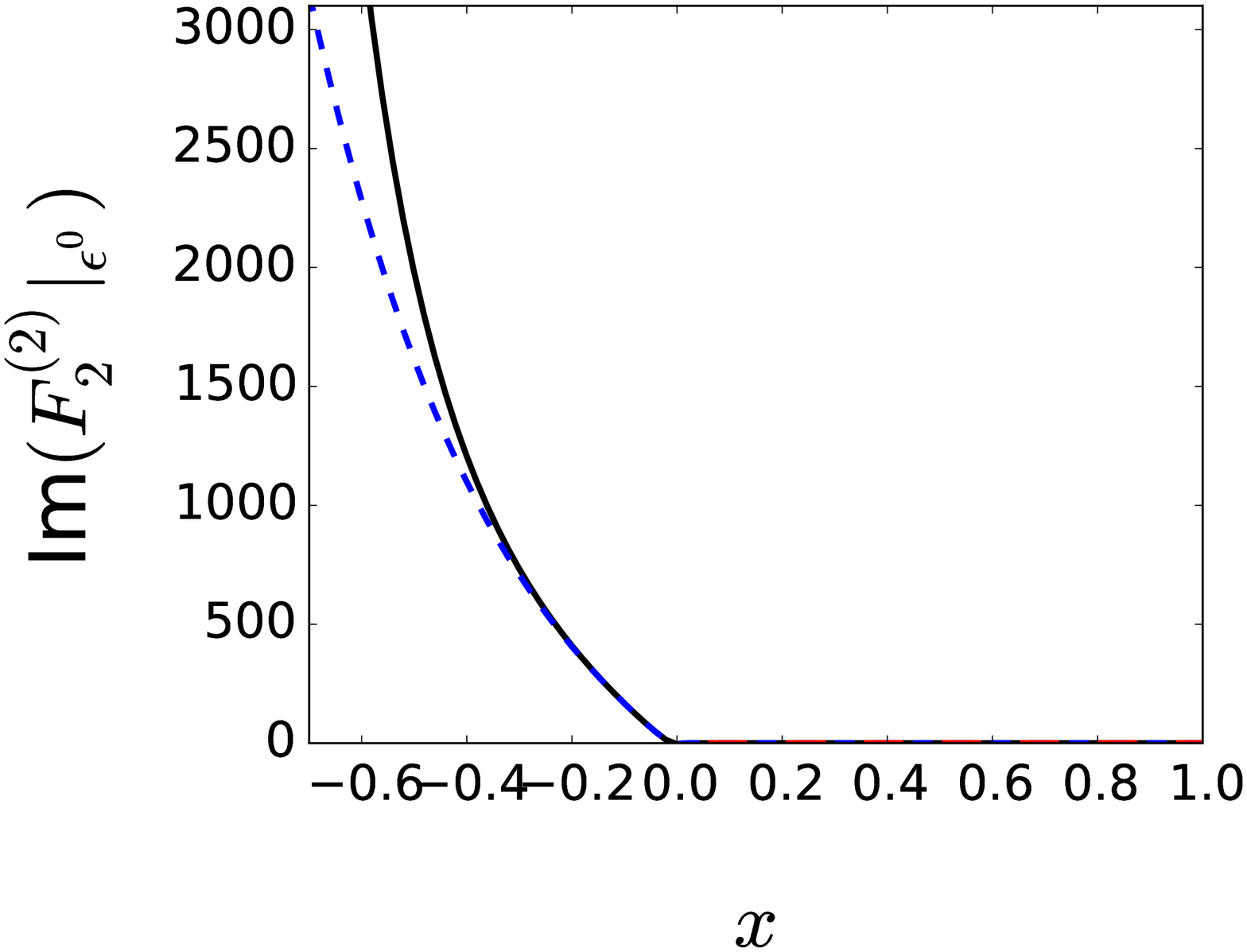}
    \includegraphics[width=.3\textwidth]{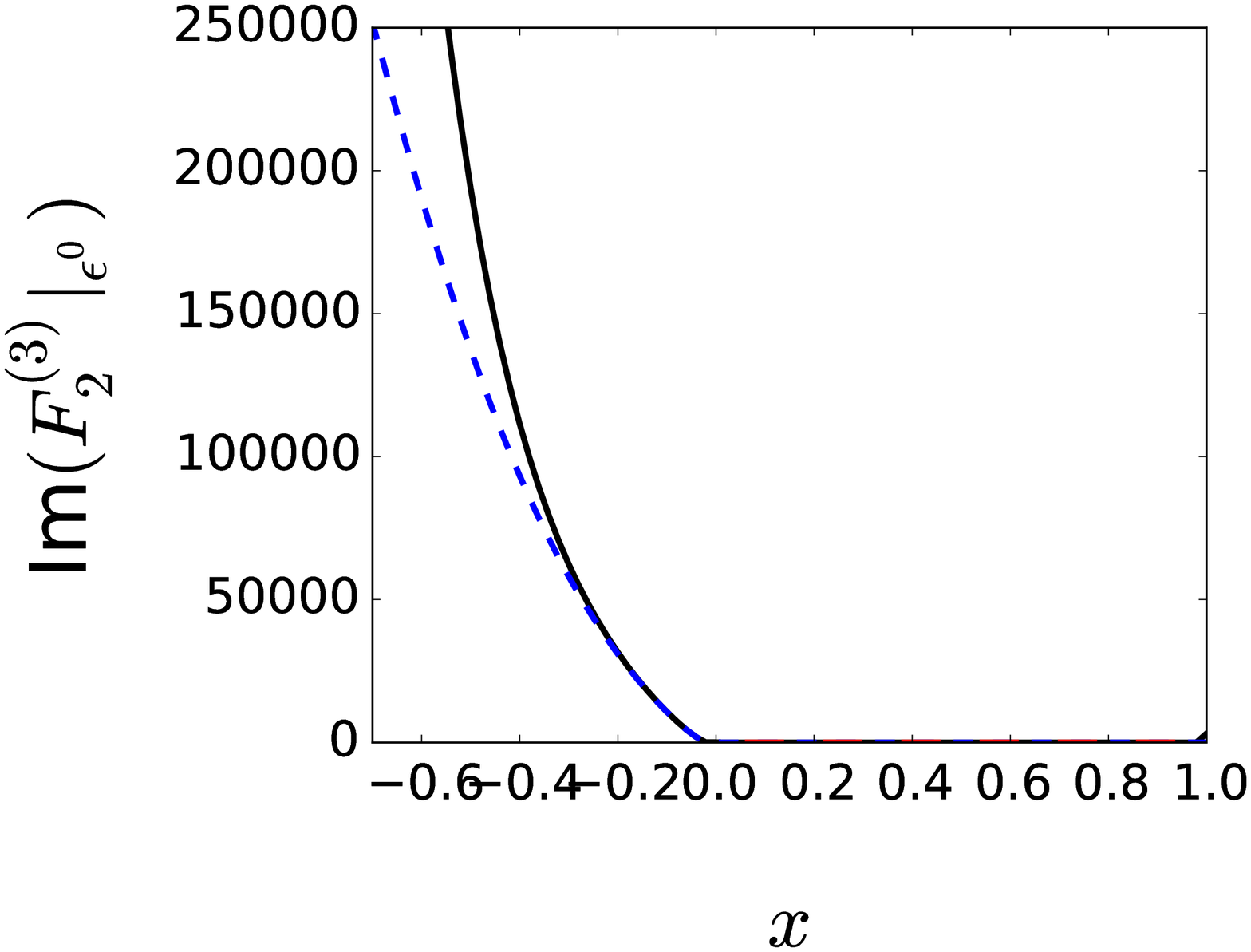}
    \caption{\label{fig::F2_x}Real and imaginary parts of $\epsilon^0$ one-,
      two- and three-loop contribution of $F_2$ as a function of $x$. The same
      notation as in Figure~\ref{fig::F1_x} has been used.}
  \end{center}
\end{figure}

This subsection is devoted to the numerical evaluation of the form factors
which we perform with the help of {\tt
  ginac}~\cite{Bauer:2000cp,Vollinga:2004sn}.  In Figs.~\ref{fig::F1_x}
and~\ref{fig::F2_x} $F_1$ and $F_2$ are shown as a function of $x$ where the
leading term of Eq.~(\ref{eq::F_i_lar}) is subtracted to obtain a regular
behaviour for $x=0$ (which corresponds to $s=\infty$). From left to right the
one-, two- and three-loop results are shown and the upper plots correspond to
the real and the lower ones to the imaginary parts.  Note that the latter are
zero for $x>0$. One observes that the expansions for $s\gg m^2$
(which include terms up to
order $x^4$) provide a good approximation to the exact result in the interval
$-0.3\lesssim x \lesssim 0.3$ which corresponds to $0.18\gtrsim m^2/s\gtrsim
-0.61$. On the other hand, the approximations obtained for
$s\ll m^2$ (which include terms up to order $(1-x)^4$) agree with the exact
result for $x \gtrsim 0.4$.

\begin{figure}[t] 
  \begin{center}
    \includegraphics[width=.3\textwidth]{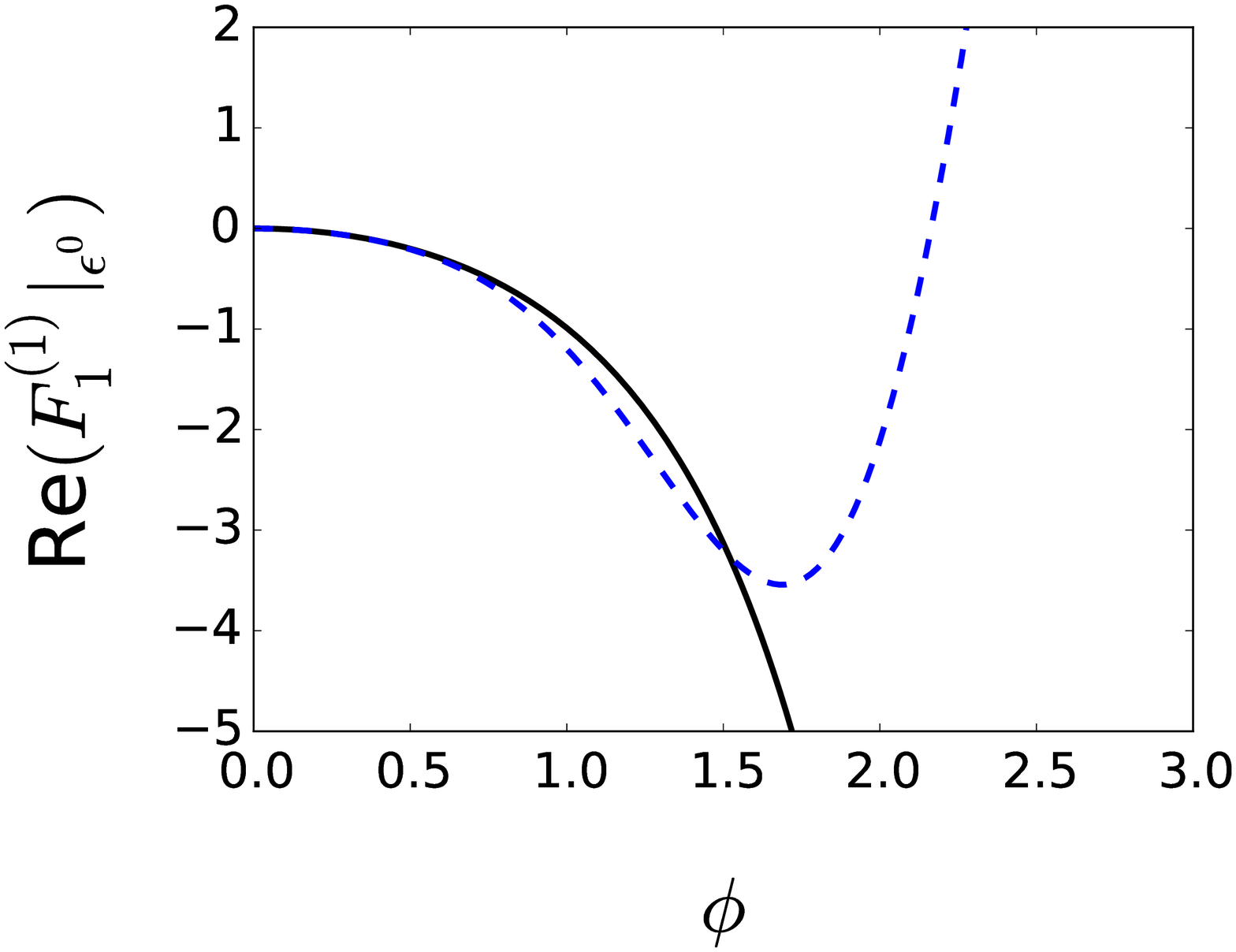}
    \includegraphics[width=.3\textwidth]{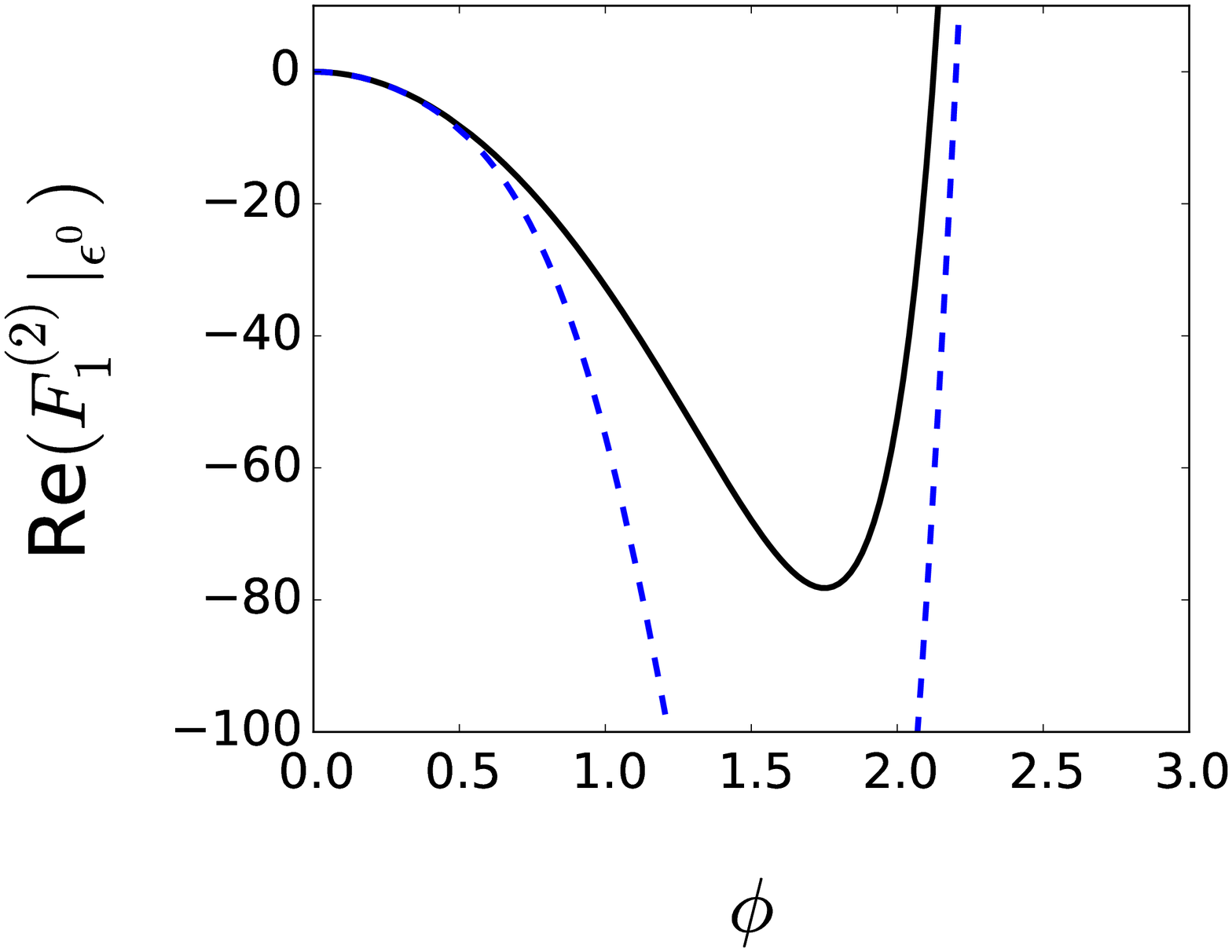}
    \includegraphics[width=.3\textwidth]{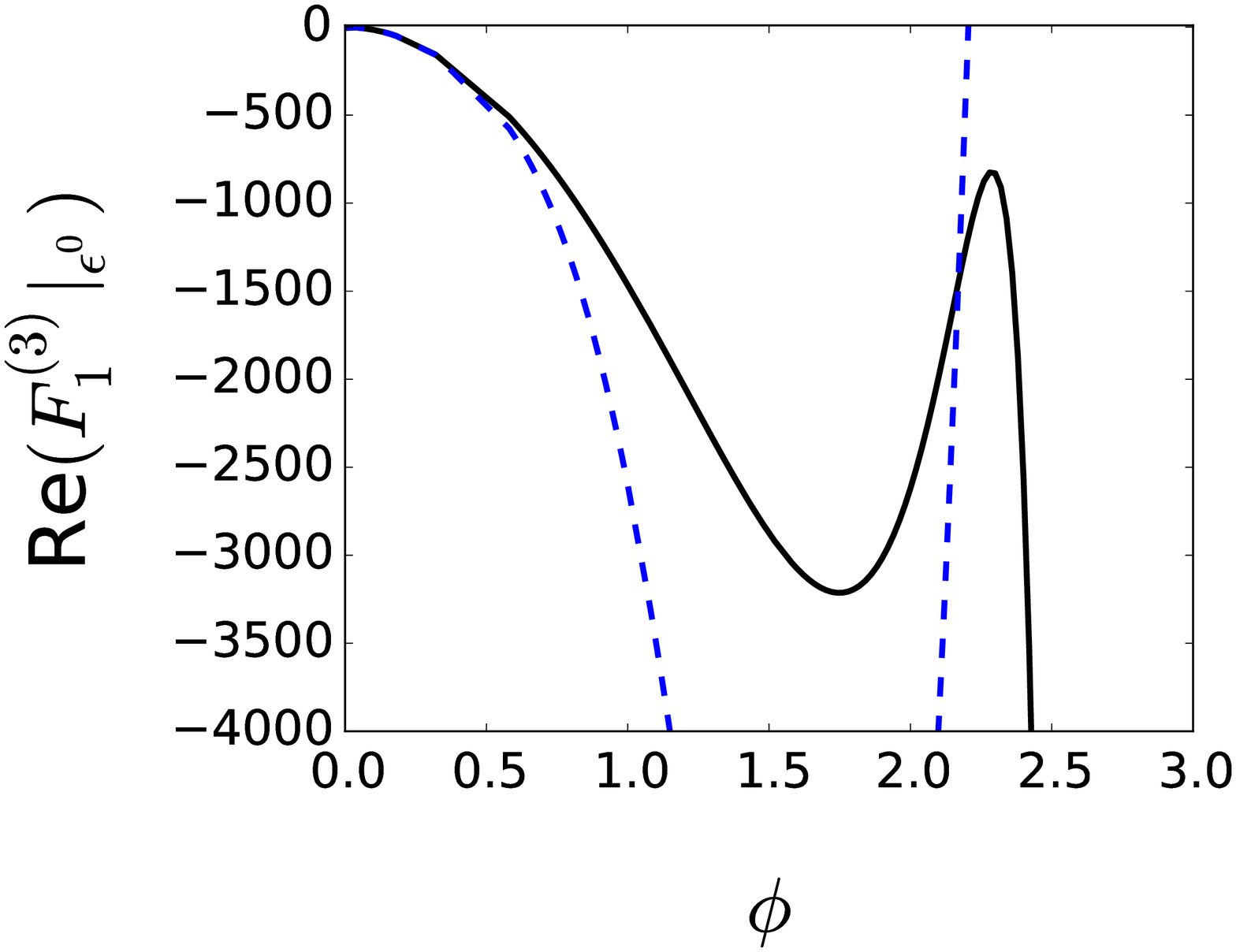}
    \\
    \includegraphics[width=.3\textwidth]{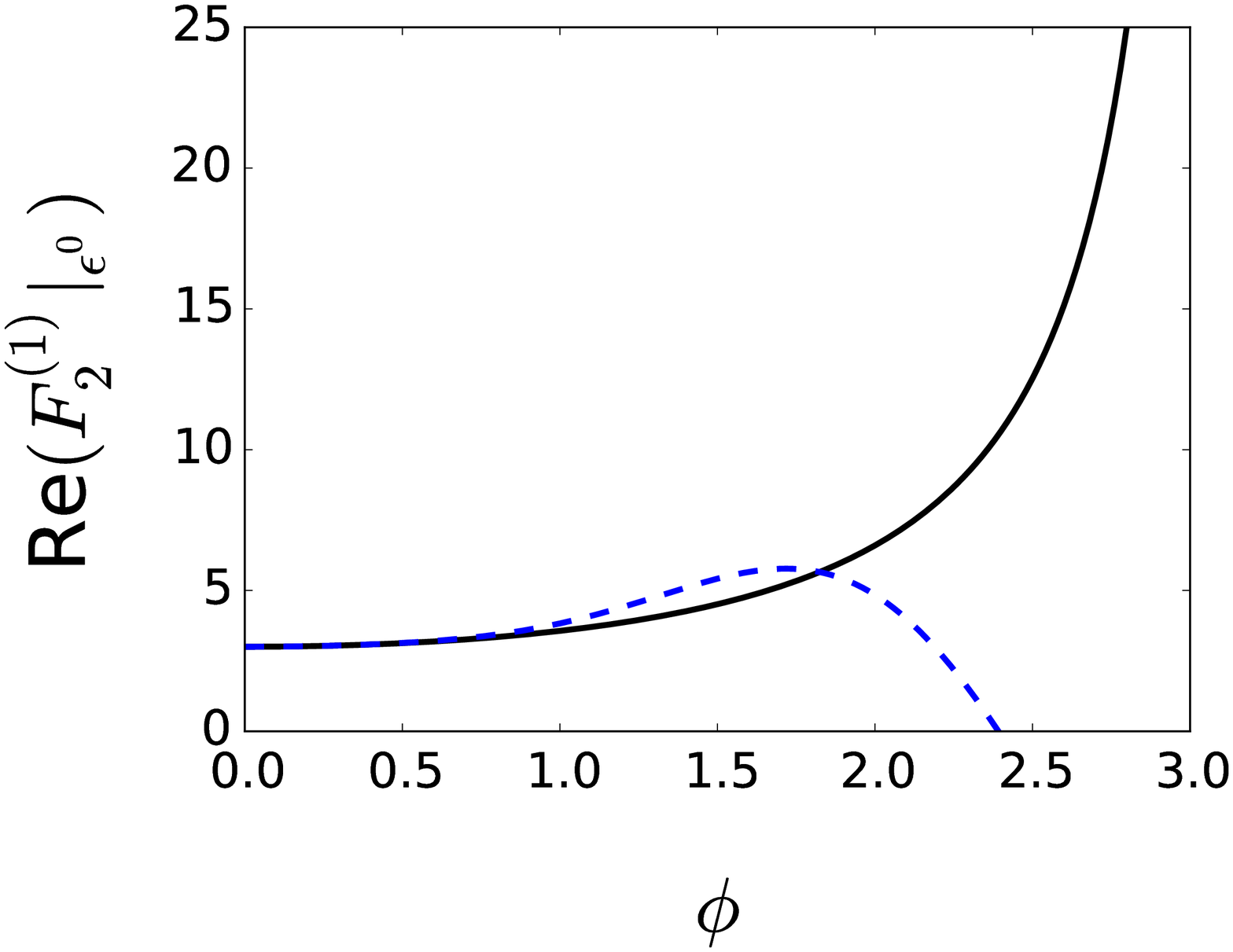}
    \includegraphics[width=.3\textwidth]{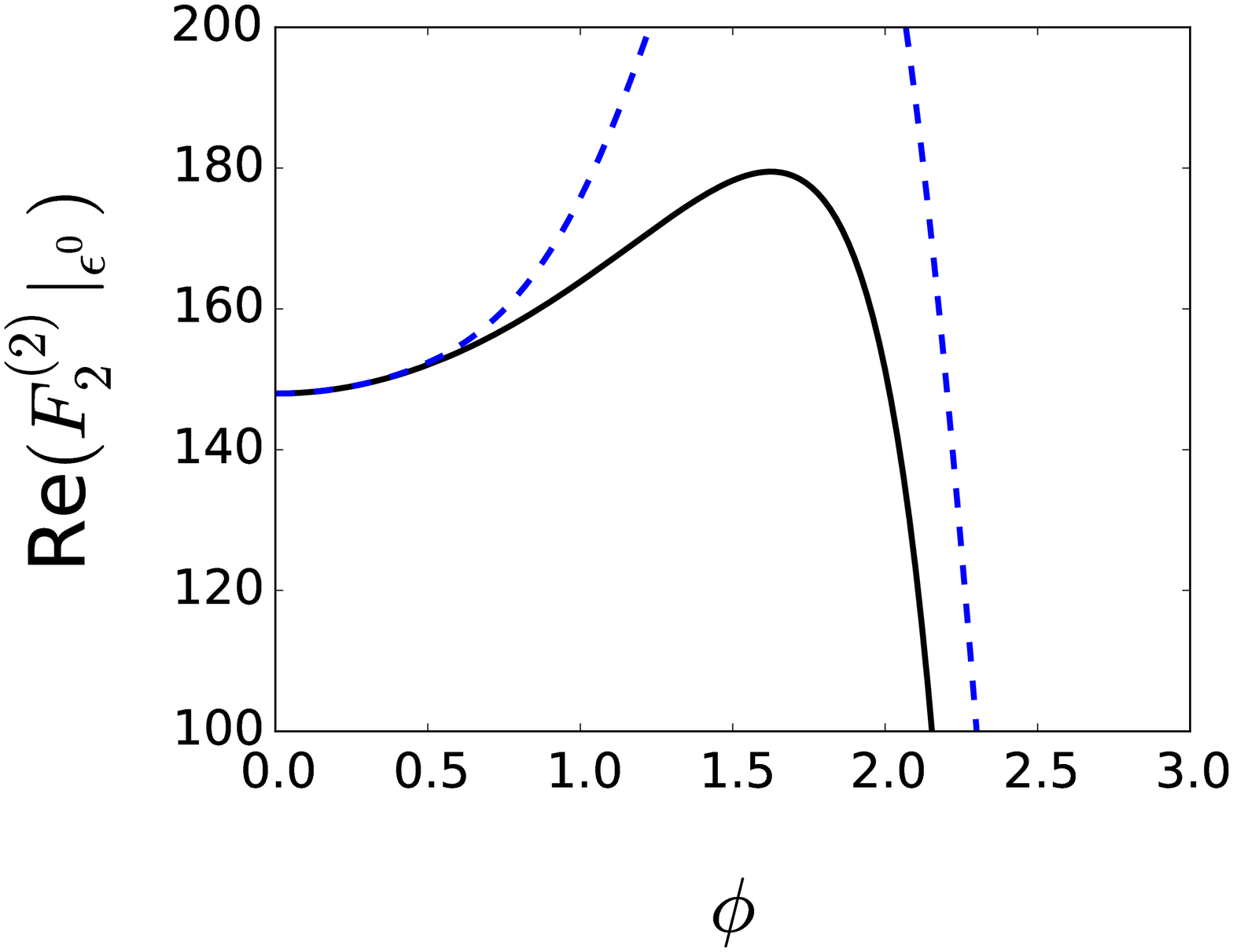}
    \includegraphics[width=.3\textwidth]{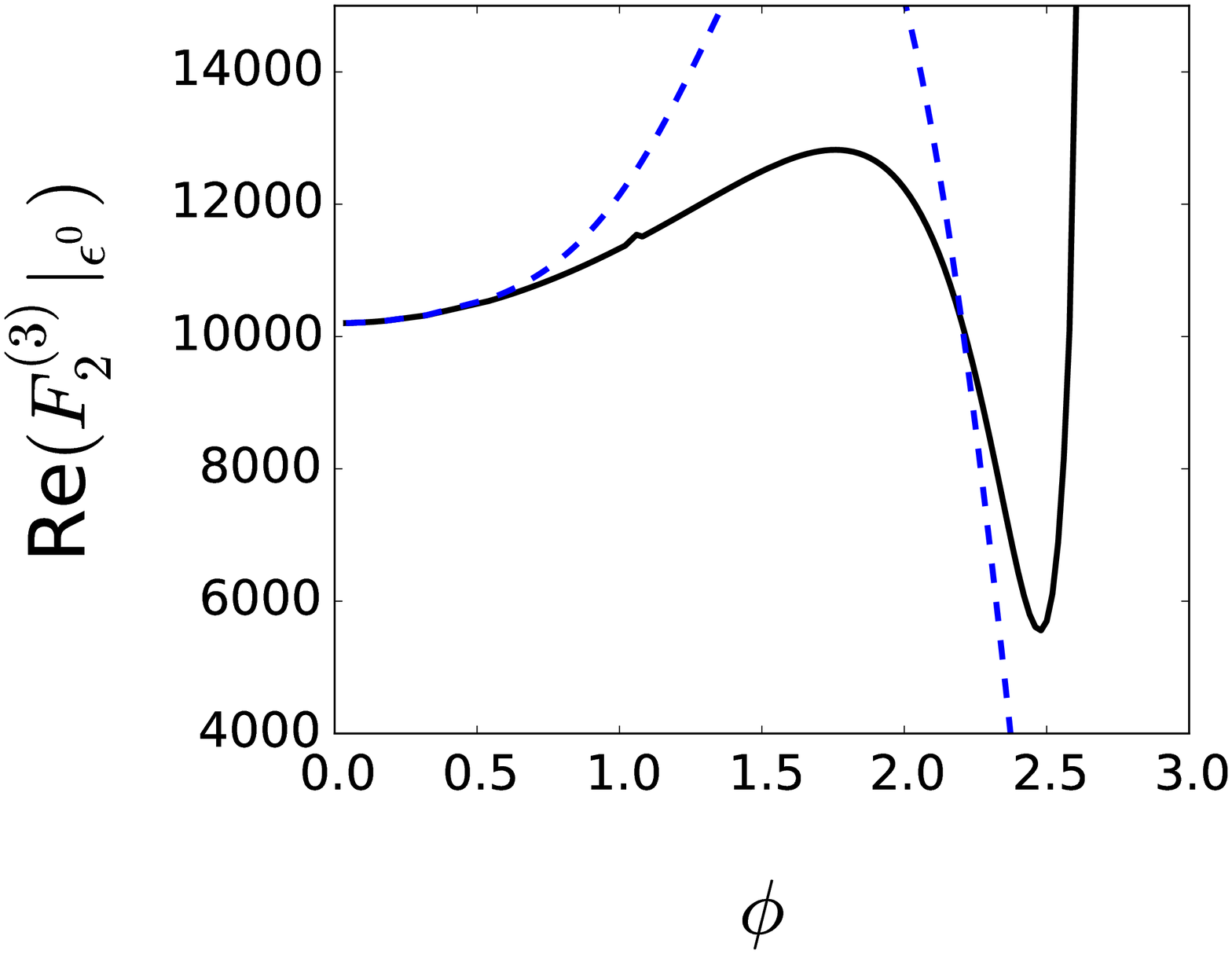}
    \caption{\label{fig::F1_F2_phi}One-, two- and three-loop contribution of
      $F_1$ ($\epsilon^0$ terms) as a function of $\phi$ (with $x=e^{i\phi}$).
      The solid (black) lines shows the exact result and the dashed (blue)
      lines represent approximations including terms up to order $(1-x)^4$.
      The number of light fermions is set to zero ($n_l=0$).  Note that in
      this region $F_1$ and $F_2$ have no imaginary parts.}
  \end{center}
\end{figure}

Fig.~\ref{fig::F1_F2_phi} shows the dependence of $F_1$ (top plots)
and $F_2$ (bottom plots) as a function of $\phi$ where $x=e^{i\phi}$.
In this region the form factors are real. One observes good agreement
of the expanded and exact result up to $\phi\approx 0.5$ which corresponds
to $s/m^2\lesssim 0.25$.

\subsection{\label{sub::checks}Checks}

Our result has passed several cross checks and consistency relations 
which we describe in this subsection.

We have successfully compared our bare and UV-renormalized
one- and two-loop results (expanded up to ${\cal O}(\epsilon^0)$) to the
expressions provided in Refs.~\cite{Bernreuther:2004ih,Gluza:2009yy} after
taking the large-$N_c$ limit.
Note that in~\cite{Bernreuther:2004ih} a different renormalization scheme has
been used which leads to a difference in the finite contribution proportional
to $\pi^2$.  This is due to the factor $\Gamma(1+\epsilon)$ which is present
in the counterterm for the strong coupling constant in Eq.~(24) of
Ref.~\cite{Bernreuther:2004ih} (see also discussion above).

For the UV-renormalized two-loop form factor $F_2$ 
we agree with Ref.~\cite{Gluza:2009yy} including ${\cal O}(\epsilon^1)$
terms. For $F_1$
we disagree with the order $\epsilon$ term of Ref.~\cite{Gluza:2009yy} in a
term which is independent of Goncharov polylogarithms. The difference of our
result and the one of~\cite{Gluza:2009yy} reads
\begin{eqnarray}
  - {\left(\frac{\alpha_s}{4\pi}\right)^2 N_c^2\,\epsilon}\,\frac{1037 x^3}{2(1 + x)^6}
  \,.
  \label{eq::diff}
\end{eqnarray}
In our expression there is no $1/(1+x)^6$ term at all.  Such a term leads to a
different low-energy and threshold behaviour. In particular, the ${\cal
  O}(\epsilon^1)$ term of the renormalized two-loop form factor would have a
stronger divergence than the expected $1/\beta^2$ behaviour, cf. the ancillary
file to this paper. Furthermore, a term as in Eq.~(\ref{eq::diff}) influences
via renormalization the $\epsilon^0$ terms of the three-loop $F_1$ which would
lead to different low-energy and threshold expansions than the ones discussed
in Section~\ref{sub::ana}. In particular, $F_1(x=1)$ would be different from
zero and the agreement of $\Delta^{(3)}$ in Eq.~(\ref{eq::Delta}) with the
literature would be destroyed.

As a further cross check we also compared with predictions
of three-loop corrections to $F_1$ in the high-energy limit which have
been obtained in Ref.~\cite{Gluza:2009yy} on the basis of evolution equations.
We find agreement including the $\log(x)/\epsilon$ terms.
The remaining $1/\epsilon$ and the $\epsilon^0$ terms
cannot be predicted using the method of Ref.~\cite{Gluza:2009yy}.
However, these terms are contained in our result.

From the $1/\epsilon$ pole of our result we can extract with the help of
Eq.~(\ref{eq::F_poles}) the cusp anomalous dimension $\Gamma_{\rm cusp}$ up
to three-loop order in the large-$N_c$ limit.  Up to two-loop order we find
agreement with Refs.~\cite{Korchemsky:1987wg,Kidonakis:2009ev} and at three
loops we can reproduce the results of~\cite{Grozin:2014hna,Grozin:2015kna}.
This is the first independent check of (part of) the results obtained
in~\cite{Grozin:2014hna,Grozin:2015kna} using a completely different method.
 
We have checked that the renormalized form factors have the correct static
limit. In particular, $F_1(0)$ vanishes and $F_2(0)$ agrees with the explicit
calculation of the three-loop corrections to the anomalous magnetic moment of
a heavy quark which was performed in Ref.~\cite{Grozin:2007fh}.

For $x\in(0,1]$ we have that $s\le0$.  Thus the results for the form factors
have to be real. Since the individual Goncharov polylogarithms are
complex-valued this is a useful cross check.

Similarly, if $x=e^{i\phi}$ with either $\phi\in[0,\pi]$ or $\phi\in[-\pi,0]$
(i.e. $x$ is on the upper or lower semi-circle) we have that $s$ is below
threshold with $0\le s/m^2 \le4$.  Again, the form factors must be
real-valued.



\section{\label{sec::concl}Conclusions and outlook}

In this paper we evaluated for the first time massive three-loop form factors, in
the planar limit.  As a byproduct, we confirmed the recent result for the
three-loop cusp anomalous dimension in the large-$N_c$ limit, which describes
the infrared divergences of the form factors.  We expressed the results
analytically in terms of Goncharov polylogarithms. The latter allow for a
straightforward numerical evaluation.

We investigated analytically the low-energy, threshold, and high-energy
limits, and derived expressions containing logarithmically enhanced as well as
power suppressed terms.  It would be interesting if some of these expansions
could be obtained from effective field theory methods. See for example
Refs.~\cite{Laenen:2010uz,Penin:2016wiw} for work on power-suppressed terms.

Our results can be used to predict infrared divergent terms at higher loop
orders, via renormalization group equations, along the lines of
Refs.~\cite{Mitov:2006xs,Gluza:2009yy}.


\section*{\label{sec::ack}Acknowledgments}
V.A.S. thanks Claude Duhr for help in manipulations with
Goncharov polylogarithms and M.S. thanks Chihaya Anzai for support
in the numerical evaluation of Goncharov polylogarithms.
We thank Taushif Ahmed and Kirill Melnikov for useful discussions.
This work is supported by the Deutsche Forschungsgemeinschaft through the
project ``Infrared and threshold effects in QCD''.  J.M.H. thanks ICTP/SAIFR
for hospitality during different stages of this work.  J.M.H. is supported in
part by a GFK fellowship and by the PRISMA cluster of excellence at Mainz
university.

\end{document}